\documentclass[useAMS,usenatbib]{mn2e} 
 
\usepackage{graphicx} 
\usepackage{txfonts} 
\usepackage{natbib}               
\newcommand\aj{AJ} 
\newcommand\apj{ApJ} 
\newcommand\apjs{ApJS}       
 
\newcommand\aap{A\&A} 
\newcommand\mnras{MNRAS} 
\newcommand\apjl{ApJ} 
\newcommand\pasp{PASP} 
\newcommand\nat{Nature}

\title[]{The \textit{HST\,} Large Program on $\omega$\,Centauri. I. Multiple stellar populations at the bottom of the main sequence probed in NIR-Optical\thanks{
    Based on observations with the NASA/ESA Hubble Space Telescope,
    obtained at the Space Telescope Science Institute, which is
    operated by AURA, Inc., under NASA contract NAS 5-26555, under
    GO-14118 and GO-14662.
}}
\author[A.\,P.\,Milone et al.] 
       {A.\,P.\,Milone$^{1}$, A.\,F.\,Marino$^{1}$, L.\,R.\,Bedin$^{2}$, J.\,Anderson$^{3}$,
         D.\,Apai$^{4,5}$, A.\,Bellini$^3$, \newauthor
          P.\,Bergeron$^{6}$, A.\,J.\,Burgasser$^{7}$, A.\,Dotter$^{8}$, J.\,M.\,Rees$^{4}$\\
$^{1}$Research School of Astronomy \& Astrophysics, Australian National University, Mt Stromlo Observatory, via Cotter Rd, Weston, ACT 2611, Australia \\
$^{2}$Istituto Nazionale di Astrofisica - Osservatorio Astronomico di Padova, Vicolo dell'Osservatorio 5, Padova, IT-35122\\
  $^{3}$Space Telescope Science Institute, 3800 San Martin Drive, Baltimore,  MD 21218, USA\\
  $^{4}$Department of Astronomy and Steward Observatory, The University of Arizona, 933 N. Cherry Avenue, Tucson, AZ 85721, USA\\
  $^{5}$Lunar and Planetary Laboratory, The University of Arizona, 1640 E. University Blvd., Tucson, AZ 85721, USA\\
  $^{6}$D\'epartement de Physique, Universit\'e de Montr\'eal,  C.P.~6128, Succ.\ Centre-Ville, Montr\'eal, Qu\'ebec H3C 3J7, Canada\\
          $^{7}$ Center for Astrophysics and Space Science, University of California San Diego, La Jolla, CA 92093, USA\\
          $^{8}$ Harvard-Smithsonian Center for Astrophysics, Cambridge, MA 02138, USA} 
\begin{document} 
\date{Draft Version Mar, 31, 2017} 
 
\pagerange{\pageref{firstpage}--\pageref{lastpage}} \pubyear{2017} 
 
\maketitle 
\label{firstpage} 

\begin{abstract}  
As part of a large investigation with {\it Hubble Space Telescope} to
study the faintest stars within the globular cluster $\omega$\,Centauri, in this work we present early results on the multiplicity of its main sequence (MS)
stars, based on deep optical and near-infrared observations.
By using  appropriate color-magnitude diagrams we  have identified, for
the first time, the two main  stellar populations I, and II along the entire MS,
from  the  turn-off  towards the  hydrogen-burning  limit.   We  have
compared the observations with suitable  synthetic spectra of MS stars
and conclude that the two  MSs are consistent with stellar populations with different metallicity, helium, and light-element abundance.
Specifically, MS-I corresponds to a metal-poor stellar population ([Fe/H]$\sim -1.7$) with Y$\sim 0.25$ and [O/Fe]$\sim$0.30.
 The MS-II hosts helium-rich (Y$\sim 0.37$-$0.40$) stars with metallicity ranging from [Fe/H]$\sim -1.7$ to $-1.4$.

Below the MS  knee ($m_{\rm F160W}\sim$19.5), our  photometry reveals  that each of  the two  main MSs hosts    stellar    subpopulations   with    different  oxygen abundances, with very O-poor stars ([O/Fe]$\sim -0.5$) populating the MS-II. Such a complexity has  never been observed in previous studies of M-dwarfs in  globular clusters. A few  months before the lunch
of the \textit{James Webb Space Telescope}, these results demonstrate the power
of  optical  and near-infrared  photometry  in  the study  of  multiple
stellar populations in globular clusters.
\end{abstract} 
 

\section{Introduction}\label{sec:intro} 
The most-massive globular cluster (GC) of the Milky Way,
$\omega$\,Centauri, hosts a very-complex system of multiple stellar
populations that makes it one of the most enigmatic stellar systems of
the Galaxy.
In contrast with the majority of mono-metallic GCs, stars in
$\omega$\,Centauri span a wide range in metallicity, including 
populations with [Fe/H]$> -0.7$ (e.g., Norris \& Da Costa\,1995;  Pancino et al.\,2002; Marino et al.\,2010; 2011; Johnson et al.\, 2008, 2009). It exhibits
extreme star-to-star variation in several light elements, including C,
N, O, Na, Mg, Al, and Si and s-process elements (e.g., Brown \& Wallerstein\,1993; Norris \& Da Costa\,1995; Johnson \& Pilachowski\,2010; Stanford et al.\,(2010); Marino et al.\,2012) and shows the typical
 abundance patterns among light elements, that are observed in nearly all the
Galactic GCs, such us the Na-O and C-N anticorrelations. Interestingly,
in this extreme cluster distinct anticorrelations are present within
stellar populations with different metallicity (Marino et al.\,2010;
2011; Johnson \& Pilachowski\,2010).
 
One of the most intriguing discoveries of the last decade in the field
of stellar astrophysics has been that the main sequence (MS) of
$\omega$\,Centauri is split into a blue and a red component (Anderson 1997; Bedin et al.\,2004) with the blue MS being also more metal rich than the red MS
(Piotto et al.\,2005). These facts have demonstrated that this cluster
exhibits a large variation in helium, with the blue MS being highly
enhanced in helium up to $Y \sim$0.39 (e.g., Norris\,2004; King et
al.\,2012). More-recent papers have shown that the MS is even more
complex than we had previously imagined and that both the blue
and the red MS host stellar subpopulations (e.g., Bellini et
al.\,2010). The complexity of multiple stellar populations in this
cluster is further demonstrated by the multiple sub-giant branches (SGBs) and red-giant branches (RGBs) and by
the presence of a double white-dwarf cooling sequence (Lee et
al.\,1999; Pancino et al.\,2000; Sollima et al.\,2005; Villanova et
al.\,2007; Bellini et al.\,2010; 2013; Milone et al.\,2017).

In this paper we use data covering one of the three parallel fields from one of the two  epochs monitored by the {\it Hubble Space
  Telescope} under Large Program GO-14118+14662 (PI: Bedin).
 A detailed description of the project and of the entire dataset is provided in the paper by Bedin et al.\,(in preparation). For this field, deep optical data exists in the archive (see Section~2 for details) and it is possible to determine cluster
membership. We exploit visual and near-infrared {\it
  HST}-photometry to investigate multiple stellar populations along
the entire MS of $\omega$\,Centauri, from the turn-off approaching the
H-burning limit.

\begin{table*}
\caption{ Column (1) provides the progressive ID of the stars analyzed in this paper. Column (2) and (3) are the absolute equatorial position tightened to Gaia-DR1 system (Lindegren et al.\,2016) in Equinox J2000.0 at epoch 2015.5; columns (4) and (5) are the coordinate positions on a reference \texttt{fits} image also released with this paper. The full version of this table is available in the electronic version of the paper.}
\begin{tabular}{cccccccccc}                                                                                    
\hline\hline
ID & RA & DEC & X & Y & F606W & F814W & F110W & F160W & member\\
(1) & (2) & (3) & (4) & (5) & (6) & (7) & (8) & (9) & (10)\\
\hline
    1 & 13:25:29.191 & $-$47:40:57.59 &  2100.66 &  1943.98  &   22.869  &   21.605  &   20.876  &   20.155  &   1 \\
    2 & 13:25:28.842 & $-$47:40:56.46 &  2151.75 &  1804.16  &   24.832  &   23.351  &   22.527  &   21.745  &   1 \\
    3 & 13:25:29.137 & $-$47:40:56.19 &  2157.97 &  1924.22  &   23.194  &   21.895  &   21.162  &   20.452  &   1 \\
    4 & 13:25:28.924 & $-$47:40:55.82 &  2176.15 &  1838.42  &   25.218  &   23.166  &   22.076  &   21.376  &   0 \\
    5 & 13:25:28.920 & $-$47:40:51.92 &  2332.99 &  1843.02  &   19.966  &   19.230  &   18.841  &   18.318  &   1 \\
\hline\hline
\end{tabular}\\
\label{tab:catalogo}
\end{table*}

The paper is organized as follows: in Section~\ref{sec:data} we describe
the data and the data reduction. The color-magnitude diagram (CMD) and its multiple sequences are presented in Sections~\ref{sec:cmd} and \ref{sec:MS}, while in Section~\ref{sec:teoria} we
compare the observed colors of the distinct stellar populations with
those predicted from appropriate synthetic spectra. A summary and a
discussion of the main results are provided in
Section~\ref{sec:conclusioni}.

\section{Data and data analysis} \label{sec:data} 
The \textit{HST} large program GO-14118+14662 aims at observing the entire
white dwarf (WD) cooling sequence of $\omega$\,Centauri. 
 It uses the \textit{Wide Field Channel} (WFC) of the \textit{Advanced Camera for Surveys} (ACS) to investigate how the unusually large helium content of some of the stars impacts the evolution of those stars as WDs. 

The field de-contamination from background and foreground objects is
particularly important when studying faint stars at sparse
evolutionary phases in star clusters.  Therefore the program was
designed to collect observations at two epochs.  While the first epoch
from GO-14118 has just concluded, the second epoch from GO-14662, will be collected between late 2017 and mid 2018.

In each of these two epochs, the 66-orbits
staring at the main field is collected at 3 different orientations, with the goal of minimizing
the impact of imperfect CTE, imperfect calibrations and artifacts, to have a better handle on systematic errors in astrometry and photometry.

As ancillary science, parallel fields with both the near-infrared
(NIR) and the UV-Visual (UVIS) channels of the \textit{Wide Field
  Camera 3} (WFC3) were also approved for a number of scientific
sub-programs. One of these scientific goals is the study of the multiple MS 
 in the vicinity of the Hydrogen burning limit.

 For each of the three orientations, the staring lasts 22 orbits.
  These are observed with both WFC3/UVIS (8 orbits) and WFC3/NIR (14 orbits).  Although these fields are not deep enough to reach the end of the WD cooling sequence, they
are deep enough to study the faintest regions of the MSs of
$\omega$\,Centauri, particularly when using WFC3/NIR where cool stars
are relatively brighter.

\subsection{The South-West Parallel Field at 17$^\prime$}

During phase II, one of the three above-mentioned parallel fields was
accurately placed to match relatively deep (GO-14118, 4 orbits)
archival observations collected in 2002 and 2005 with ACS/WFC
(more details will be presented in the following sections).
%
For completeness, this field was not chosen as the main ACS/WFC field due to its relatively low density of cluster stars. 
This field is centered at $(\alpha;\delta)\simeq($13:25:36.8$;-$47:40:05$)$, i.e., about 17 arc-minutes from the cluster center.
 This field is the focus of this paper.

In this work we use  \textit{only} the WFC3/NIR images from GO-14118 
 collected between August 19
and August 26 2015, and consisting of 7 short exposures of
 142\,s and 14 long exposures of 1302\,s for each of the
 two filters F110W and F160W.  The adopted exposure times have been chosen to maximize the dynamical range and optimize the detection of faint sources\footnote{Specifically, the short exposures were collected with a number of sampling equal to 15 (\texttt{NSAMP=15}), organized in linear sampling mode with steps of 10\,s (\texttt{SAMP-SEQ=SPARS10}), while deep exposure with \texttt{NSAMP=14} in mode \texttt{SAMP-SEQ=SPARS100}. See Wide Field Camera 3 Instrument Handbook (for Cycle 25) http://www.stsci.edu/hst/wfc3/documents/handbooks/currentIHB/\-wfc3\_cover.html, Sect~7.7.3 MULTIACCUM Timing Sequences: Full Array Apertures for details.}.

Data were reduced with an adaptation to WFC3/NIR images of the software
presented and described in Anderson et al.\ (2008) for
the case of the ACS/WFC camera.
The software is essentially a PSF-fitting algorithm that performs
simultaneous fitting of all the sources in a given patch of sky, 
 by simultaneously using all the available images for that patch of
sky, once transformed into a common reference frame. To achieve this, the software makes use
of the the publicly available PSFs and geometric-distortion solution for
WFC3/NIR provided by Jay Anderson.\footnote{\texttt{http://www.stsci.edu/$\sim$jayander/WFC3/}
}

\subsection{Archival Material}

The parallel field considered in this work was previously observed
with ACS/WFC on July 7th 2002 under program GO-9444, and this is the
field in which the MS split of $\omega$\,Centauri was first demonstrated (Anderson 1997; Bedin et al.\,2004).
The data consists of 2-orbits per filter, F606W and F814W, for a total of four $\sim$1300\,s deep exposures per filter.  
A second  epoch was awarded to obtain high-precision proper-motion membership (GO-10101), and data were collected on December 25$^{\rm th}$ 2005.
Details for these fields were presented in King et al.\,(2012). 
We independently re-reduced the two data-sets from GO-9444 and GO-10101, using the software described in Anderson et al.\,(2008).
We then averaged the stellar magnitudes observed in the two data-sets.

 GO-9444 data were taken with an instrumental setup (\texttt{GAIN=1}) which does not allow us to recover the charge along the bleeding columns. Indeed, in that setup the digital saturation occurs before the physical one and the excess charge is irremediably lost. 
In contrast, the photometry of bright saturated stars were obtained from GO-10101
 images, which were collected with an instrumental setup that allows 
 the charge excess to be recovered (\texttt{GAIN=2}), as demonstrated by Gilliland (2004). For saturated stars, we used the positions and the fluxes provided by King et al.\,(2012) and derived by fitting the wings of the PSFs (see Section~8.1 from Anderson et al.\,2008 for details).

 \begin{centering} 
 \begin{figure*}
  \includegraphics[width=12.5cm]{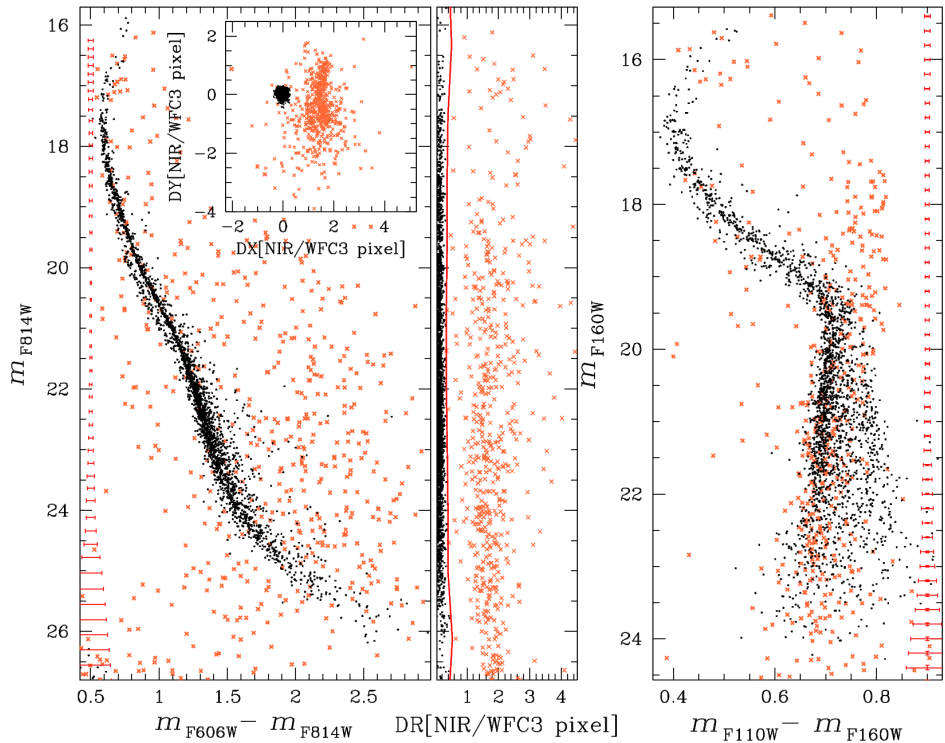}
  \caption{$m_{\rm F814W}$ vs.\,$m_{\rm F606W}-m_{\rm F814W}$ (left
    panel) and $m_{\rm F160W}$ vs.\,$m_{\rm F110W}-m_{\rm F160W}$
    (right panel) CMD of stars in the NIR/WFC3 field for which
    proper-motion measurements are available. The inset in the left
    panel shows the vector-point diagram of the relative stellar
    displacements. Middle panel shows $m_{\rm F814W}$ as a function of
    the total displacement DR. The red line separates the probable
    cluster members from field stars, which are represented with black
    dots and orange crosses, respectively, in all the panels of the
    figure.
  }
  \label{fig:cmdvpd} 
 \end{figure*} 
 \end{centering} 

\subsection{Calibration, proper motions, and differential reddening}\label{sub:DR}
 Photometry has been calibrated into the Vega magnitude system by following the recipe of Bedin et al.\,(2005). For WFC/ACS photometry we have used the zero points from Bedin et al.\,(2005), while in the case of NIR photometry we have adopted the zero points provided by STScI web page for WFC3/NIR\footnote{http://www.stsci.edu/hst/wfc3/phot\_{zp}\_{lbn}}.

 Since the multiple MSs are the main target of this paper, we have analyzed only relatively isolated stars, which are well fitted by the PSF model and have small magnitude and position rms. These stars have been selected by using the criteria described in detail by Milone et al.\,(2009) and Bedin et al.\,(2009).
  Briefly, for each star we have calculated a number of parameters that can be used as diagnostics of the photometric and astrometric quality. These diagnostics include: the photometric and astrometric rms, the fraction of light in the aperture due to neighbors (o), the quality of the PSF fit (q), and the excess of flux outside the PSF core (RADX). We plotted each parameter against the stellar magnitude and verified that most of the stars define a clear relation as a function of the magnitude. Outliers are stars with poor astrometic and photometric quality and are not considered in this paper.
 Photometry has been corrected for differential reddening by using the procedure described by Milone et al.\,(2012).  To estimate the amount of differential reddening associated to each star, we first derived the fiducial line of MS-I stars in the $m_{\rm F160W}$ vs.\,$m_{\rm F606W}-m_{\rm F160W}$ plane, which is the most sensitive CMD to differential reddening. Then, we calculated the residuals between the color of each star and the color of the fiducial along the reddening line. We assume as best estimate of the differential reddening associated to each star, the median of the residual values of the 75 neighbours MS-I stars. We refer to Section~3.1 from Milone et al.\,(2012) for further details.
 
  We have used the relative proper motions of the stars in the analyzed field of view to separate field stars from cluster members. Proper motions have been derived as in previous papers from our group (e.g.\,Anderson \& King 2003; Bedin et al.\,2003) by comparing the distortion-corrected coordinates of stars derived from the GO-14118 and GO-9444 data sets, which provide the longest time base-line. 
 We make public the astrometric and photometric catalogue of all sources studied in this paper. The first five lines of the catalogue are listed in Table~\ref{tab:catalogo}. We also release an astrometrized stacked image in the band F814W. Cols (6)--(9) are
the calibrated magnitudes into the Vega-mag system $m_{\rm  F606W}$, $m_{\rm F814W}$, $m_{\rm F110W}$, $m_{\rm F160W}$ (where the
magnitude is not available for a given filter the value is flagged to \texttt{99.9999}); finally in Col (10) we give a flag for
proper-motion based membership: 1 for members, and 0 for field objects.

\section{The color-magnitude diagram}\label{sec:cmd}
 \begin{centering} 
 \begin{figure*} 
  \includegraphics[width=12.5cm]{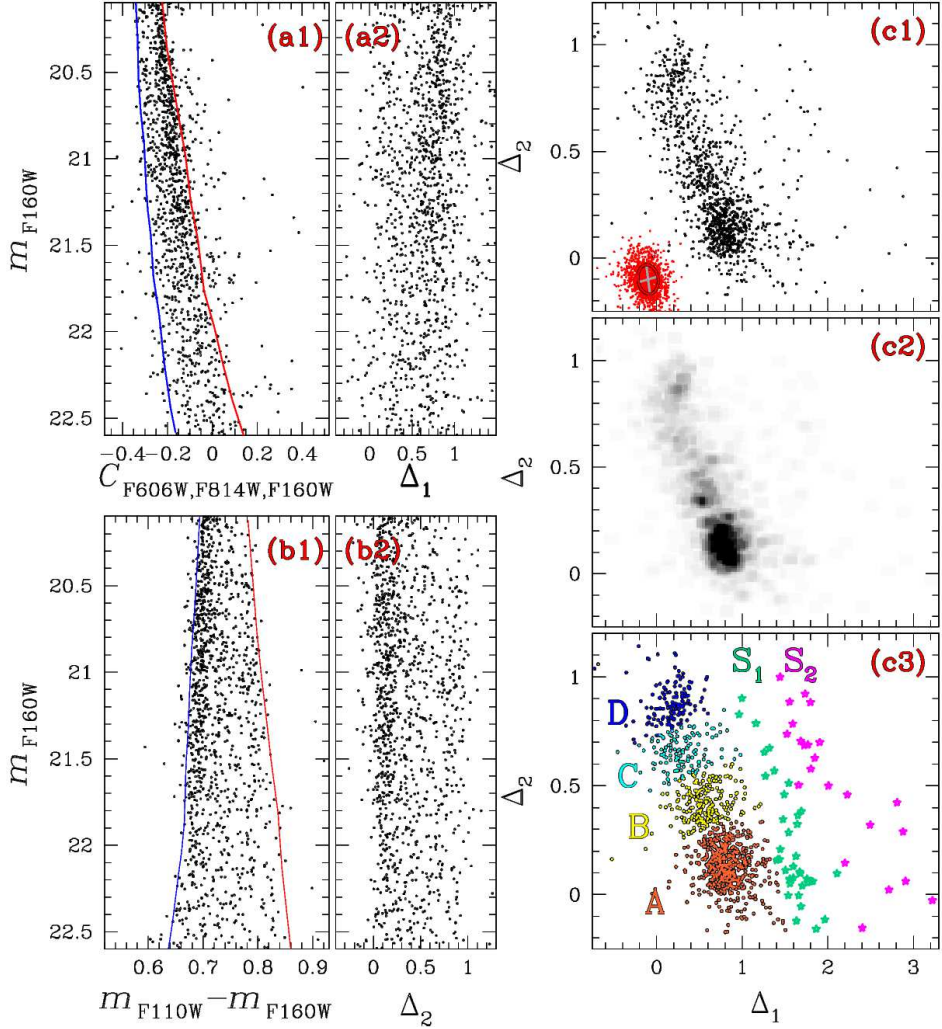} 
  \caption{Panels a1 and b1 show the $m_{\rm F160W}$ vs.\,$C_{\rm F606W, F814W, F160W}$, and $m_{\rm F160W}$ vs.\,$m_{\rm F110W}-m_{\rm F160W}$ diagram, respectively, for MS stars fainter than the MS knee. The red and the blue line super-imposed on each diagram represent the fiducial lines used to derive the corresponding verticalized $m_{\rm F160W}$ vs.\,$\Delta_{1}$ and $m_{\rm F160W}$ vs.\,$\Delta_{2}$ diagrams plotted in panels a2 and b2. The $\Delta_{2}$ vs.\,$\Delta_{1}$ pseudo two-color diagrams or `chromosome map' is plotted in panel c1 and c3, while panel c2 shows the corresponding Hess diagram. The red points indicate the stellar distribution expected from observational errors only, where the red ellipses includes the 68.27\% of points. In the chromosome map in panel c3 we have represented stars of populations A, B, C, D, S$_{1}$, and S$_{2}$ with orange, yellow, cyan, blue, aqua, and magenta symbols, respectively (see text for details). } 
  \label{fig:mappe} 
 \end{figure*} 
 \end{centering} 
The left panel of Fig.~\ref{fig:cmdvpd} shows the $m_{\rm F814W}$
vs.\,$m_{\rm F606W}-m_{\rm F814W}$ CMD for all the stars that pass
the criteria of selection discussed in Section~\ref{sub:DR} and
 for which both proper motions and NIR photometry are available. The
inset shows the vector-point diagram of the stellar displacements, DX
and DY, in units of NIR/WFC3 pixel, while the F814W magnitude is
plotted as a function of the total displacement DR=$\sqrt{{\rm DX}^{2}+{\rm DY}^{2}}$ in the middle panel of
Fig.~\ref{fig:cmdvpd}.  The red line separates candidate cluster
members (black points) from field stars (orange crosses). To
derive this boundary, we have divided the magnitude range with
 $15.75<m_{\rm  F814W}<26.75$ into intervals of 0.25 mag and for each of these we 
have derived the mean magnitude and the median and the $68^{\rm th}$
percentile of the DR displacements of the cluster members (hereafter
$\sigma_{\rm DR}$). The red line was derived by linearly interpolating the
 points with abscissa equal to four times $\sigma_{\rm DR}$ and ordinate corresponding to the mean magnitudes.  The sample of cluster members used to derive the
red line has been determined iteratively. At the first iteration we
  used all the stars with DR$<$0.4 pixel, while at the subsequent
iterations we considered as candidate cluster members all the
stars on the left side of the red line. The procedure has been
repeated three times until convergence.

A visual inspection of the CMD of cluster members plotted in the left
panel of Fig.~\ref{fig:cmdvpd} reveals that the blue and the red MSs of
$\omega$\,Centauri,  discovered by  Bedin et  al.\,(2004), are  clearly
visible
from the brightest magnitude before saturation down to 
an interval of about three magnitudes in the F814W band.  At fainter
luminosity, $m_{\rm F814W} \gtrsim 21.4$, the two distinct blue and the red
MS are not distinguishable and most of the cluster stars form a single
and broadened MS.  The only exception is provided by a third
poorly-populated red MS, previously identified by King et al.\,(2012),
 which includes only a few percent of the total number of MS stars.
  
The NIR, $m_{\rm F160W}$ vs.\,$m_{\rm F110W}-m_{\rm F160W}$, CMD is
plotted in the right panel of Fig.~\ref{fig:cmdvpd}. In this case the
blue and the red MS are visible in the upper part of the CMD and merge
together at the level of the MS knee around $m_{\rm F160W} \sim 19.5$. A
double MS is clearly visible at fainter magnitudes, where the
more-populated MS has bluer $m_{\rm F110W}-m_{\rm F160W}$ colors in contrast with what we observe above the MS knee, where the blue MS is less populated.
Noticeably, below $m_{\rm F160W} \sim 19.5$, most of the field stars
follow the more-populated MS of $\omega$\,Centauri, suggesting peculiar properties for the stars in the less-populated component.
 
\section{The complex MS of $\omega$\,Centauri}\label{sec:MS}
In this section we identify the multiple stellar populations along the MS.
We first analyze the region of the CMD below the MS knee in Section~\ref{subsec:bottom}, while in Section~\ref{subsec:2pop} we identify and investigate the two main stellar populations of $\omega$\,Centauri along the entire CMD, from the SGB towards the hydrogen-burning limit.

\subsection{Multiple populations at the bottom of the MS}
\label{subsec:bottom} 
In order to identify multiple populations of M-dwarf stars we start by extending the procedure introduced by Milone et al.\,(2015) in their study of the GC NGC\,7089 (M\,2) to MS stars with $20.1<m_{\rm F160W}<22.6$. 

The  procedure by Milone and collaborators in summarized in the following and illustrated in Fig.~\ref{fig:mappe}. Panel a1 shows the $m_{\rm F160W}$ vs.\,$C_{\rm F606W, F814W, F160W}$=($m_{\rm F606W}-m_{\rm F814W}$)$-$($m_{\rm F814W}-m_{\rm F160W}$) pseudo CMD of stars in the analyzed interval of magnitude. The red and the blue lines are the blue and the red envelopes of the MS and have been derived as follows.
 We have first divided the F160W magnitude interval in a series of magnitude bins of width $\nu=0.25$ mag by following the naive estimator method (Silverman 1986). The bins are defined over a grid of $N$ points separated by steps of fixed magnitude ($s=\nu/5$). For each bin, i, we have calculated the  5$^{\rm th}$ and the 95$^{\rm th}$ percentile of the $m_{\rm F606W}-m_{\rm F160W}$ color distribution ($x1_{\rm i}$ and $x2_{\rm i}$) and associated these points with mean $m_{\rm F160W}$ magnitude of the stars in the bin ($y_{\rm i}$). The points ($x1_{\rm i}, y_{\rm i}$) and  ($x2_{\rm i}, y_{\rm i}$) were then smoothed to obtain the blue and the red line shown in panel a1 of Fig.~\ref{fig:mappe}, respectively.
 We have used the boxcar averaging smoothing, where each point has been replaced by the average of the three adjacent points.   

We verticalized the CMD of the stars shown in panel a1 by using the relation:
\begin{equation}\label{eq:1}
\Delta_{1}= [(X-X_{\rm red~fiducial})/(X_{\rm blue~fiducial}-X_{\rm red~fiducial})]-1 
\end{equation}
where $X=C_{\rm F606W, F814W, F160W}$. We show $m_{\rm F160W}$ vs.\,$\Delta_{1}$ in the panel a2 of Fig.~\ref{fig:mappe}.

We derived the quantity $\Delta_{2}$  similarly, but by using the \,$m_{\rm F160W}$ vs.\,$m_{\rm F110W}-m_{\rm F160W}$ diagram shown in the panel b1 of Fig.~\ref{fig:mappe}. The magnitude $m_{\rm F160W}$ is plotted against $\Delta_{2}$ in panels b1.
In panel c1 of Fig.~\ref{fig:mappe} we show the $\Delta_{2}$ vs.\,$\Delta_{1}$,  diagram for the analyzed MS stars of $\omega$\,Centauri, while panel c2 shows the corresponding Hess diagram. This diagram is similar to the one defined by Milone et al.\,(2015, 2017) and will be indicated as `chromosome map' by following the nickname by Renzini et al.\,(2015). All the panels of Fig.~\ref{fig:mappe} show the same stars.

The red points plotted in the bottom-left corner of panel c1 represent the distribution of stars that we would expect from observational errors only  and are randomly extracted from a Gaussian distribution where the dispersion corresponds to the uncertainties. The fact that the observed $\Delta_{1}$ and $\Delta_{2}$ distribution of stars is significantly wider than the corresponding error distribution proves that the chromosome map of $\omega$\,Centauri is not consistent with a simple population.
 A visual inspection at the figures of panels c1 and c2 suggests that four main bumps are present along this sequence that we indicate as A, B, C, and D and color orange, yellow, cyan, and blue, respectively, in the $\Delta_{2}$ vs.\,$\Delta_{1}$ diagram shown in panel c3. 
 
 This diagram reveals two additional poorly-populated sequences of stars that span the interval of $-0.2\lesssim \Delta_{2}\lesssim 1.0$ and we name them S$_{1}$ and S$_{2}$. They are represented with aqua and magenta starred symbols, respectively and, as we will see in Sect.~\ref{subsec:2pop}, are well separated in the $m_{\rm F814W}$ vs.\,$m_{\rm F606W}-m_{\rm F160W}$ CMD.
 
\begin{centering} 
 \begin{figure*} 
  \includegraphics[width=12.5cm]{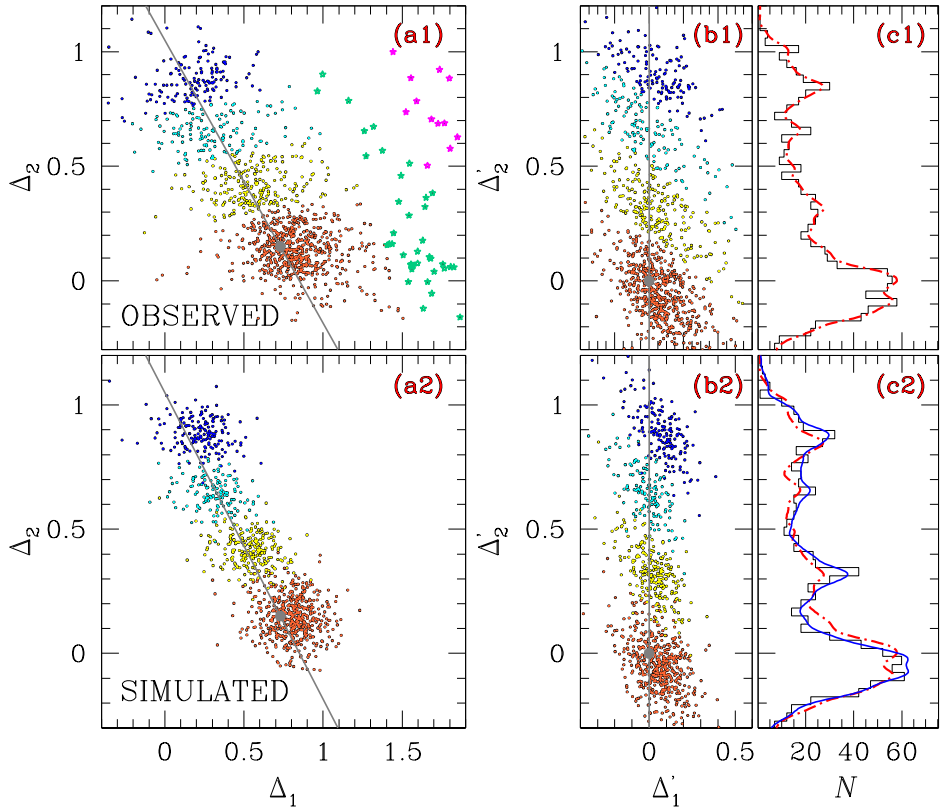} 
  \caption{ Reproduction of the chromosome map for the observed $\omega$\,Centauri stars plotted in panel c3 of Fig.~\ref{fig:mappe} (panel a1) and simulated chromosome map of four stellar populations (panel a2). Panels b1 and b2 show the corresponding verticalized $\Delta^{'}_{2}$ vs.\,$\Delta^{'}_{1}$ diagrams, while the  histogram distribution of the $\Delta^{'}_{2}$ quantities for observed population A--D and simulated stars are plotted in panels c1 and c2, respectively. The $\Delta^{'}_{2}$ kernel-density distribution of the observed stars is represented with red dashed-dotted lines in both panel c1 and c2, while the blue line shown in panel c2 marks the corresponding distribution for simulated stars (see text for details).} 
  \label{fig:mappe2} 
 \end{figure*} 
 \end{centering}

 To demonstrate that the selected A--D stellar groups are consistent with four stellar populations we compare in Fig.~\ref{fig:mappe2} the observed chromosome map of $\omega$ Centauri stars (panel a1) and the chromosome map of four simple stellar populations. We named these populations A$_{\rm SIM}$, B$_{\rm SIM}$, C$_{\rm SIM}$, and D$_{\rm SIM}$ and colored orange, yellow, cyan and blue, respectively, in panel a2 of Fig.~\ref{fig:mappe2}.
We assumed that all the simulated stars in the population A$_{\rm SIM}$ are clustered around the average $\Delta_{1}$ and $\Delta_{2}$ values of the observed populations-A stars and  adopted similar criteria for the other three groups of simulated stars. The $\Delta_{1}$ and $\Delta_{2}$ scatter around the average value that is visible in the panel a2 of Fig.~\ref{fig:mappe2} is due to observational errors.  Simulations shown in panel a2 are obtained by randomly extracting points from a two-dimensional Gaussian distribution, where the dispersion corresponds to the observed uncertainties. 

To compare the observed and the simulated stars, we first arbitrarily selected the gray dot and draw the straight line shown in panels a1 and a2 of Fig.~\ref{fig:mappe2}. Than we translated and rotated all the stars of panels a1 and a2 in such a way that the origin of the new reference frame corresponds to the gray dot and the gray line translates into a vertical line with ordinate equal to zero. The derived $\Delta^{'}_{2}$ vs.\,$\Delta^{'}_{1}$ diagram is plotted in panels b1 and b2 for observed and simulated stars, respectively.

Panel (c1) of Fig.~\ref{fig:mappe2} shows the $\Delta^{'}_{2}$ histogram diagram for observed populations A--D stars while the corresponding kernel-density distribution, derived by using a Gaussian kernel with dispersion of 0.02 mag, is represented with the red dashed-dotted line.
The $\Delta^{'}_{2}$ histogram distribution for simulated stars is plotted in panel (c2). Both the histogram distributions and the kernel-density distribution for simulated (blue line) and observed stars (red dashed-dotted) line are quite similar. This fact demonstrates that the four groups A--D of stars are actually consistent with four stellar populations.

Similarly, we note that the $\Delta_{1}$ and $\Delta_{2}$ distribution of $S_{1}$ and $S_{2}$ stars is significantly wider than what we expect from observational errors only thus indicating that each sequence hosts more than one stellar population.

A visual inspection at the chromosome maps shown in panels a1 and a2 reveals that the  populations A--D are more-clearly separated in the simulated diagram than in the observed ones. This fact is quite expected. Indeed, it is well known from literature studies on the chromosome map of $\omega$\,Centauri RGB stars that this cluster exhibit at least fifteen stellar populations (Milone et al.\,2017, see also Bellini et al.\,2010; Marino et al.\,2011; Villanova et al.\,2014). This suggests that each group of population A--D may host sub-populations of stars with  different chemical composition that contribute to the $\Delta_{2}$ vs.\,$\Delta_{1}$ broadening of each population.

 \subsection{The two main stellar populations of $\omega$\,Centauri}
 \label{subsec:2pop}

\begin{centering} 
 \begin{figure*} 
  \includegraphics[width=15.5cm]{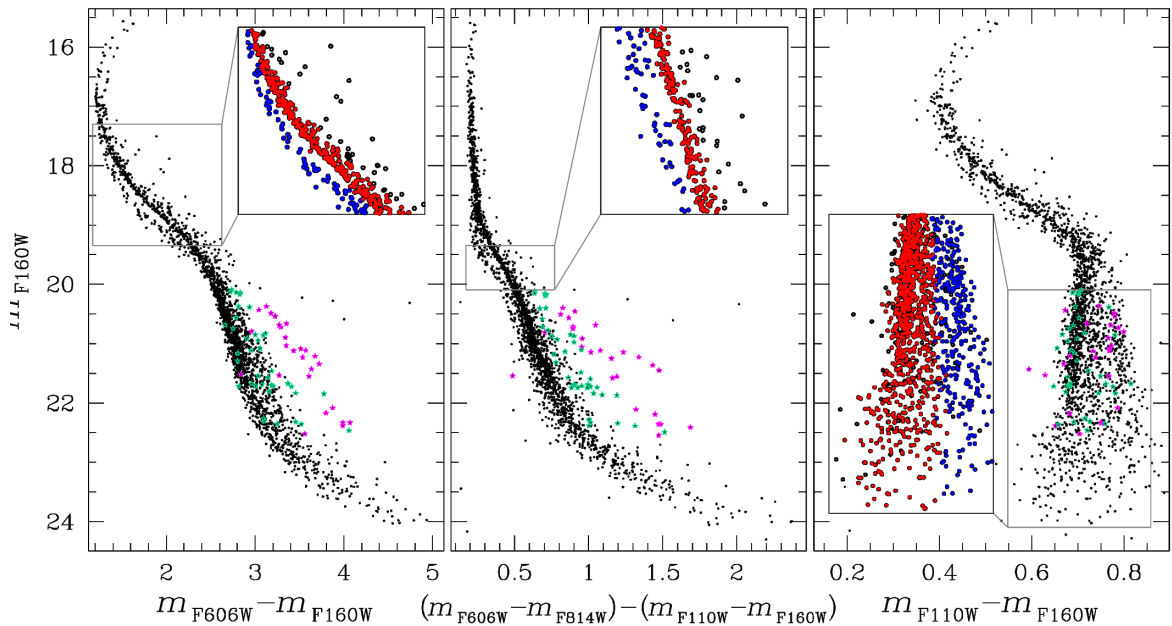} 
  \caption{$m_{\rm F160W}$ vs.\,$m_{\rm F606W}-m_{\rm F160W}$ CMD (left), $m_{\rm F160W}$ vs.\,($m_{\rm F606W}-m_{\rm F814W}$)$-$($m_{\rm F110W}-m_{\rm F160W}$) (middle), and $m_{\rm F160W}$ vs.\,$m_{\rm F110W}-m_{\rm F160W}$ CMD  (right) of $\omega$\,Centauri stars. The insets are zoom of the corresponding diagrams where the two MS are clearly visible. Red and blue colors indicate MS-I and MS-II stars, respectively. S$_{1}$ and S$_{2}$ stars identified in Fig.~\ref{fig:mappe} are colored aqua and magenta, respectively.
  } 
  \label{fig:cmdBMS} 
 \end{figure*} 
 \end{centering}

To further investigate the multiple stellar populations along the MS, we show in  
Fig.~\ref{fig:cmdBMS} two CMDs and a pseudo CMD\footnote{ We indicate as pseudo CMD a diagram, like ($m_{\rm  F606W}-m_{\rm F814W}$)$-$($m_{\rm F110W}-m_{\rm F160W}$), where the ordinate corresponds to the stellar magnitude and the abscissa to a color difference.} of
$\omega$\,Centauri stars obtained from different combinations of colors
and magnitudes.
In each diagram the two most-evident MSs are well
separated only at specific and well-defined magnitude intervals. Specifically, in the $m_{\rm F160W}$ vs.\,$m_{\rm F606W}-m_{\rm F160W}$ and the $m_{\rm  F160W}$ vs.\,$m_{\rm F110W}-m_{\rm F160W}$ CMDs the two main
populations are clearly visible in the upper and the lower MS,
respectively. In the $m_{\rm F160W}$ vs.\,($m_{\rm  F606W}-m_{\rm F814W}$)$-$($m_{\rm F110W}-m_{\rm F160W}$) pseudo CMD, the split MS is visible at intermediate luminosity only.

 In order to identify the two main populations of $\omega$\,Centauri along the entire MS, we have combined information from the three diagrams of Fig.~\ref{fig:cmdBMS}.
 We have first selected by eye from the left-panel CMD a sample of MS-I and MS-II stars in the magnitude range $17.30\leq m_{\rm F160W}<19.35$ where the two populations are clearly separated. Similarly, the middle-panel pseudo-CMD has been used to select MS-I and MS-II stars with  $19.35\leq m_{\rm F160W}<20.10$.

 The separation between the two main MSs is less evident in the NIR CMD plotted in the right-panel of Fig.~\ref{fig:cmdBMS}. Nevertheless, the majority of the stars populate the bluest MS component which is clearly connected with MS-I stars identified in the upper MS. Conversely, the remaining MS stars are distributed towards the red.
 
 We note that in the magnitude range $20.10\leq m_{\rm F160W}<22.60$ the most-populated MS hosts most of the population-A and -B stars identified in Fig.~\ref{fig:mappe},    while population-C and -D stars mostly correspond to the other MS component.
  In this magnitude interval we will thus associate the sample of population-A- and B-stars to the MS-I, and population-C- and D-stars to the MS-II.
  At fainter magnitudes we consider as MS-I and MS-II stars all the stars with $m_{\rm F110W}-m_{\rm F160W}$ greater than or smaller than 0.38 mag, respectively.
  The selected MS-I and MS-II stars are colored blue and red, respectively, in the insets of Fig.~\ref{fig:cmdBMS}.

 \begin{centering} 
 \begin{figure*} 
   \includegraphics[height=8.5cm]{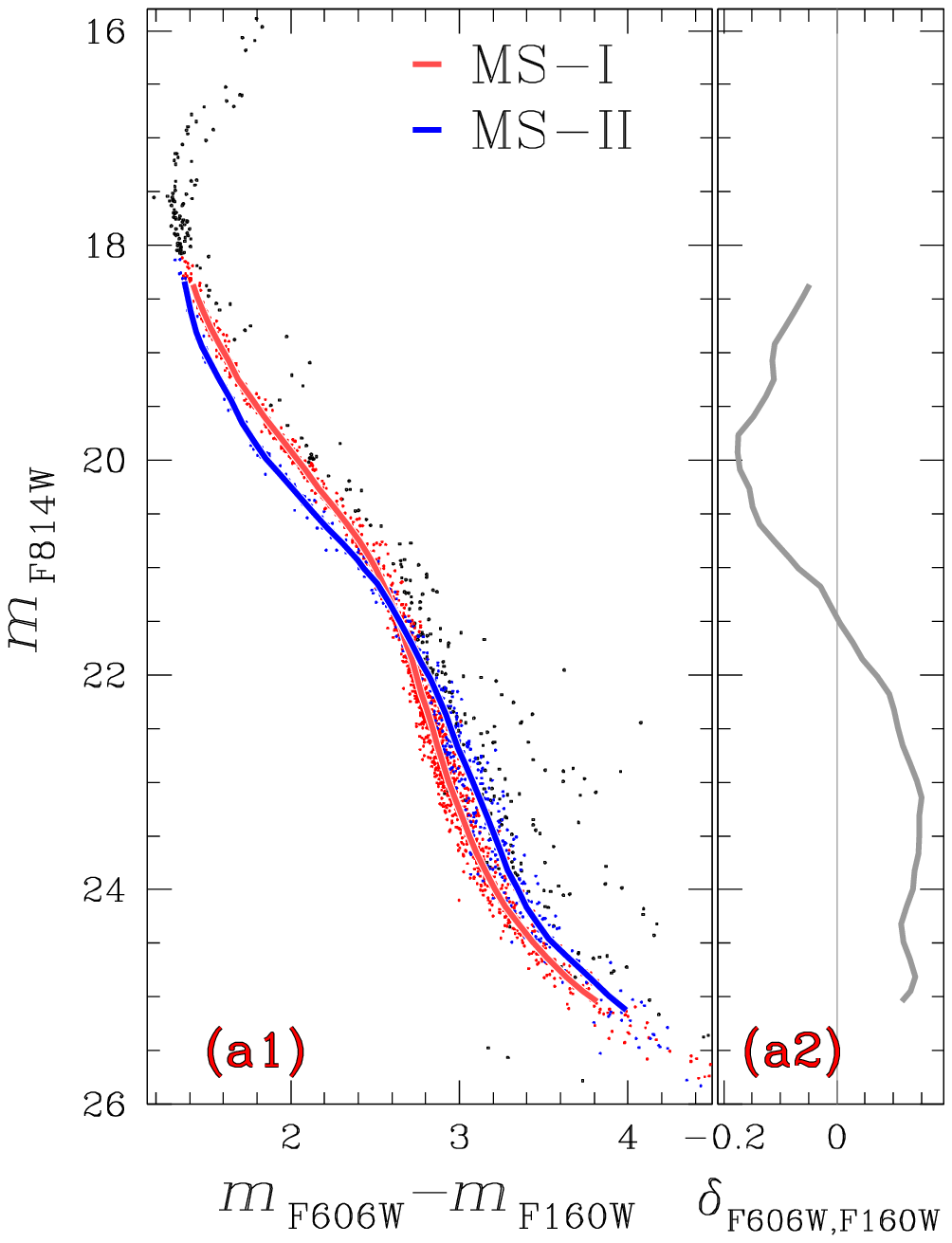}
   \includegraphics[height=8.5cm]{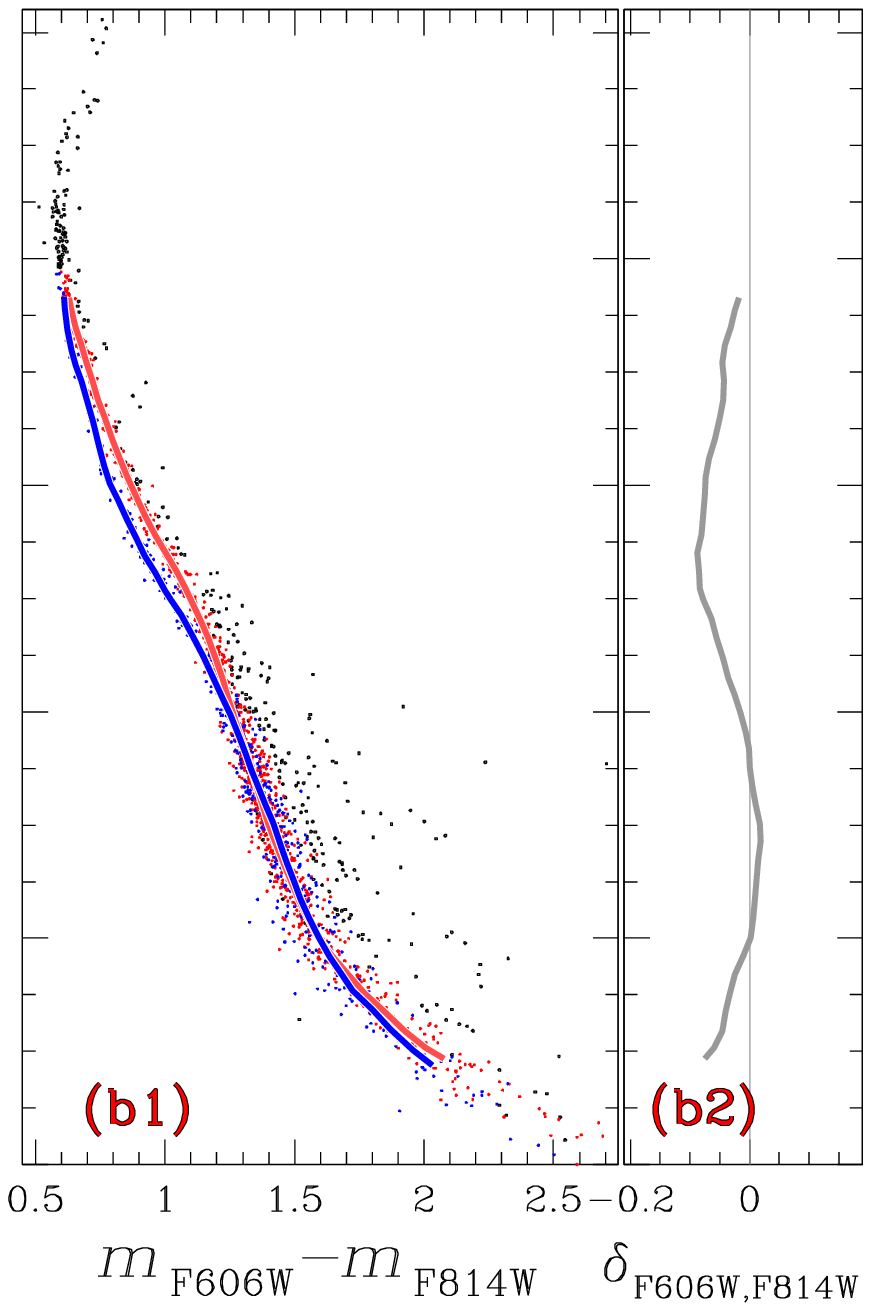}
   \includegraphics[height=8.5cm]{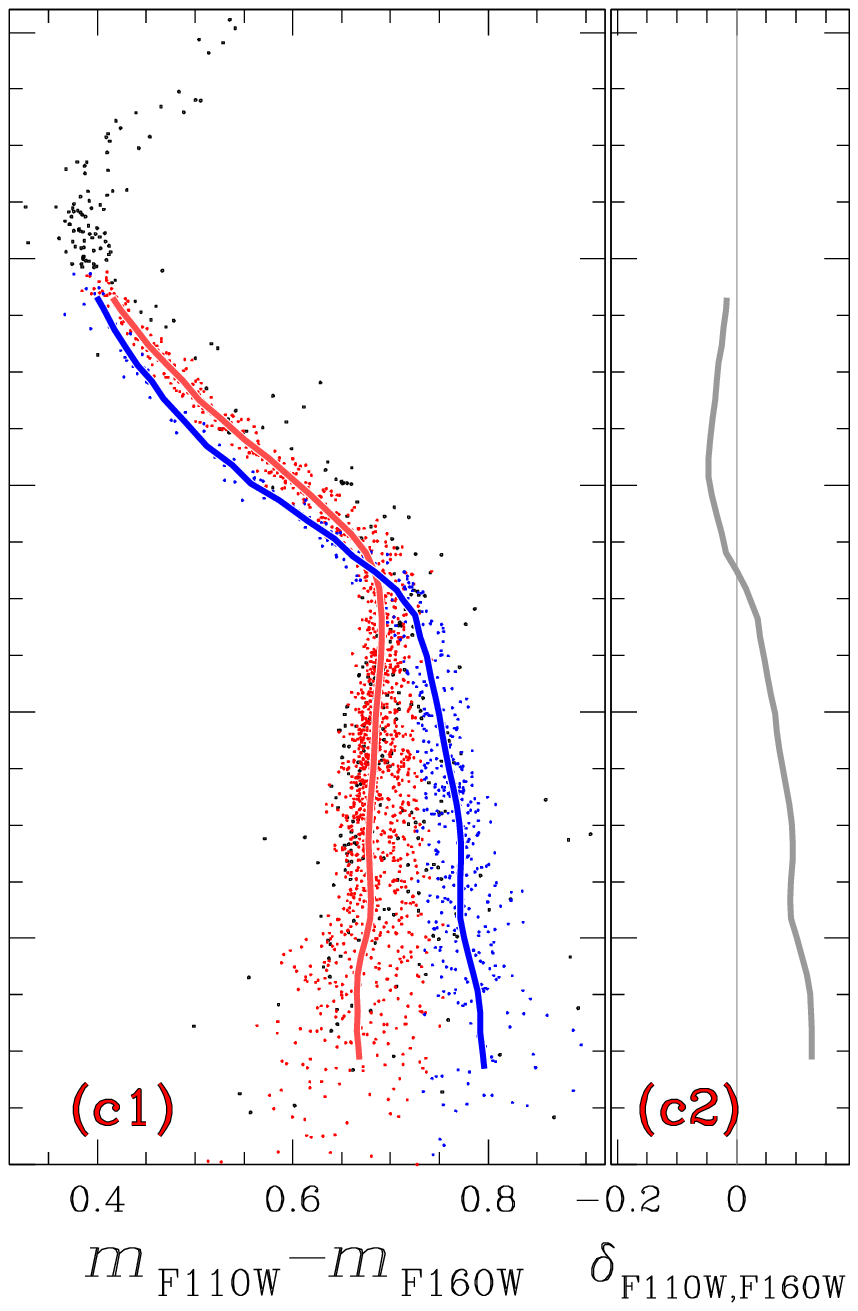} 
  \caption{Collection of optical and near-infrared CMD of $\omega$\,Centauri of Fig.~\ref{fig:cmdvpd}. The sample of MS-I and MS-II stars identified in Fig.~\ref{fig:cmdBMS} are colored red and blue in the panels a1, b1, and c1, respectively, while the continuous lines are the corresponding fiducial lines. In the panels a2, b2, and c2 we have represented with thick gray lines the F814W magnitude as a function of the color difference between the blue and red fiducial (see text for details).} 
  \label{fig:MSRLs} 
 \end{figure*} 
 \end{centering} 

 \begin{centering} 
 \begin{figure*} 
   \includegraphics[height=8.5cm]{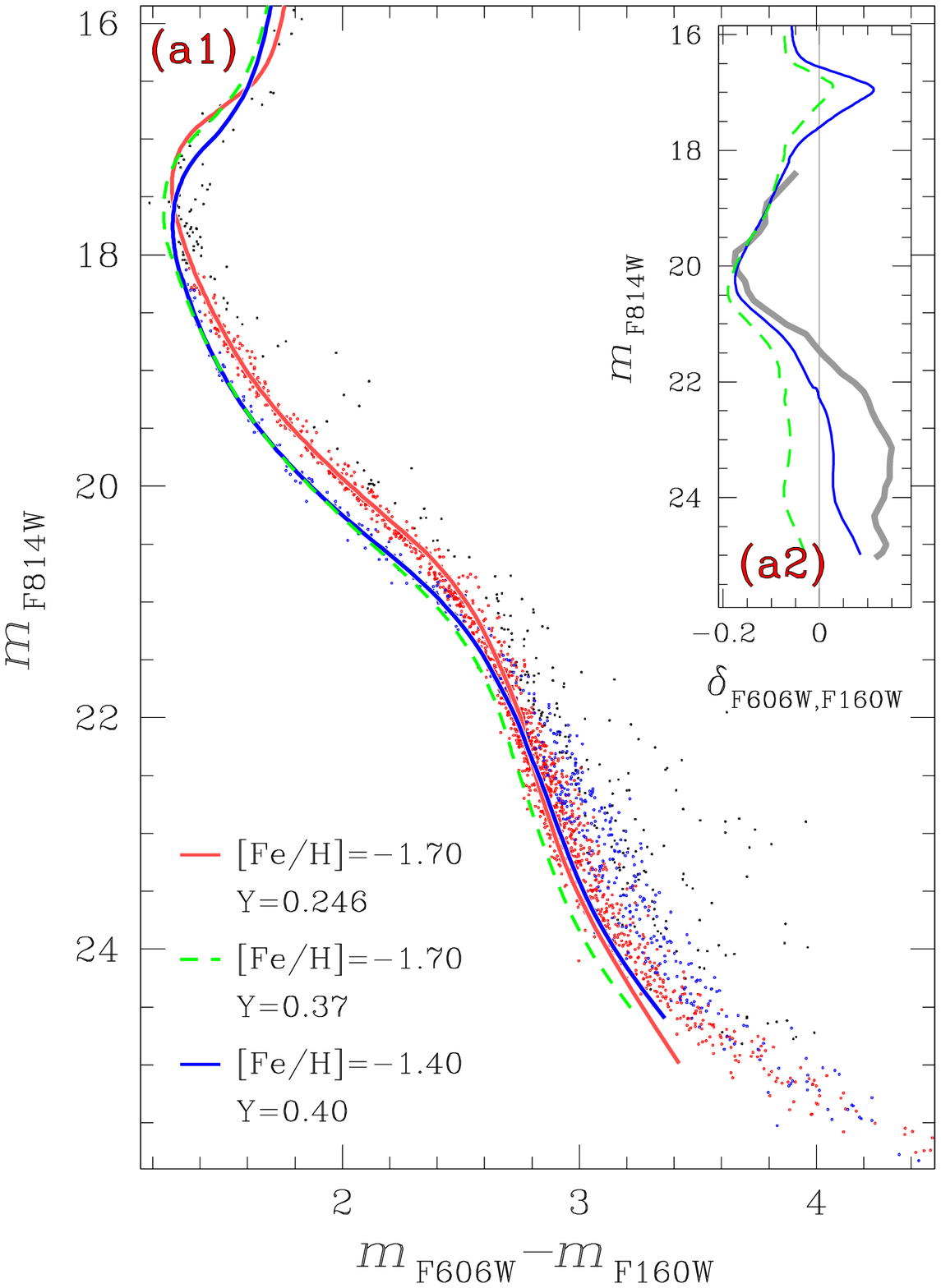}
   \includegraphics[height=8.5cm]{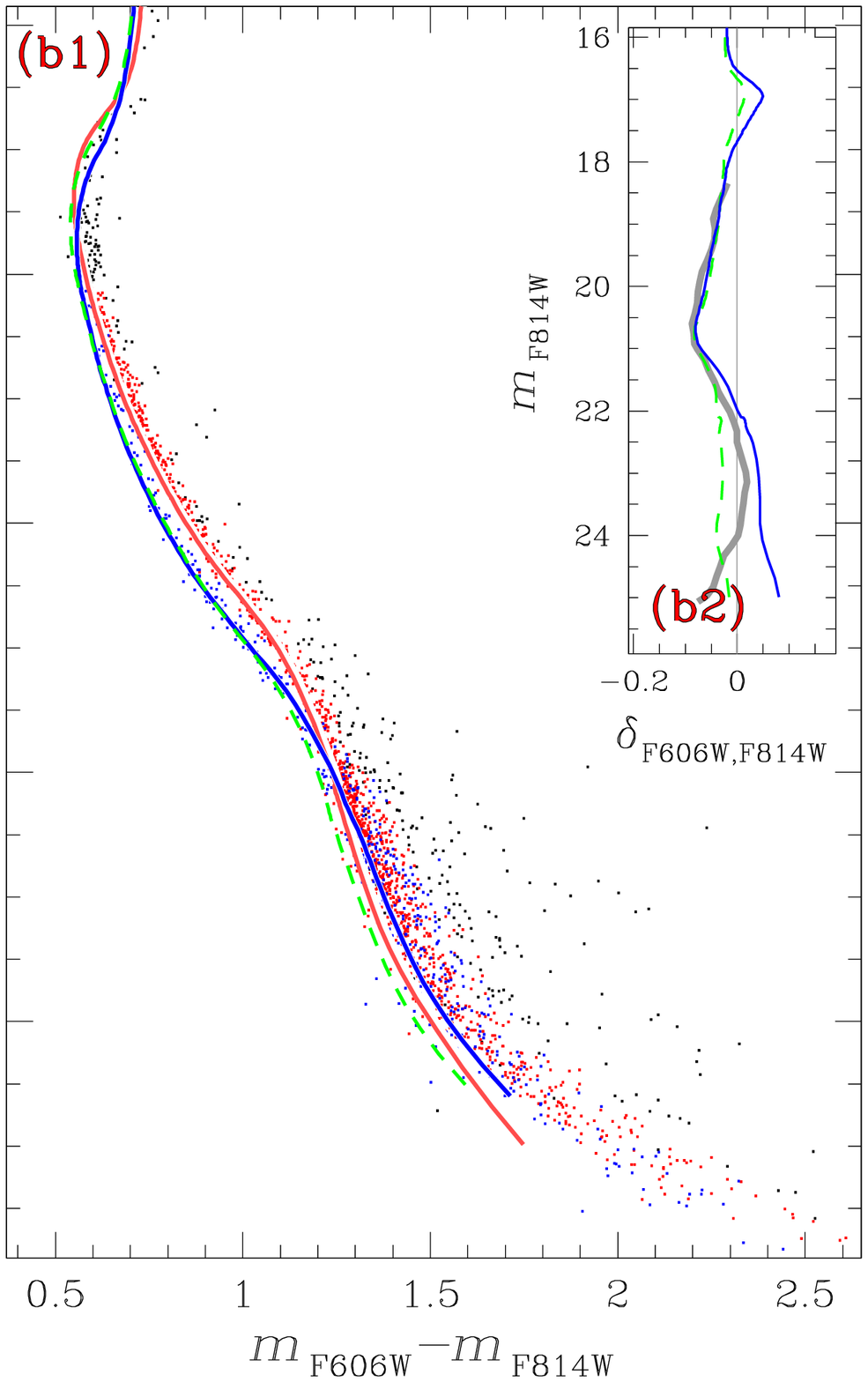}
   \includegraphics[height=8.5cm]{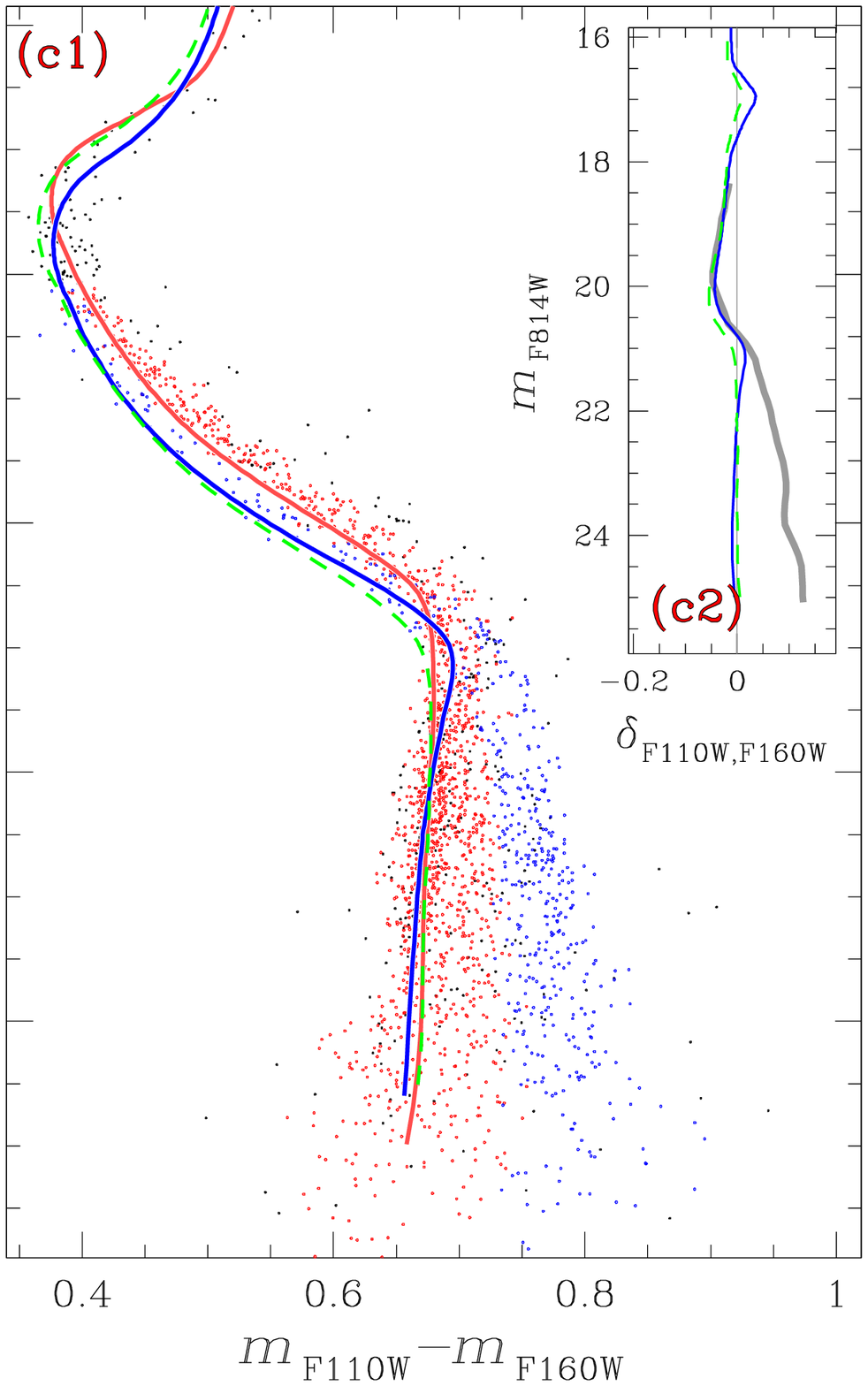} 
  \caption{Comparison between the observed CMDs of Fig.~\ref{fig:MSRLs} and isochrones from Dotter et al.\,(2008). MS-I and MS-II stars are colored red and blue, respectively. The red isochrones corresponds to the stellar population with primordial helium abundance and [Fe/H]=$-1.70$.  The blue-continuous and green-dashed isochrones correspond to the helium-enhanced stellar populations with [Fe/H]=$-1.40$ and [Fe/H]=$-1.70$, respectively (see text for details).  In the insets a2, b2, and c2 we have compared the color difference between each isochrone and the metal-poor isochrone and the corresponding color difference between the fiducial of the MS-II and that of the MS-I.} 
  \label{fig:iso} 
 \end{figure*} 
 \end{centering} 
 
 The $m_{\rm F160W}$ vs.\,$m_{\rm F606W}-m_{\rm F160W}$  CMD shown in the bottom-left panel of Fig.~\ref{fig:cmdBMS} reveals that the MS-I and the MS-II clearly correspond to the red and the blue MS discovered by Anderson\,(1997) and Bedin et al.\,(2004). A third, poorly populated MS that exhibits redder $m_{\rm F606W}-m_{\rm F160W}$ and $m_{\rm F814W}-m_{\rm F160W}$ colors than the majority of MS stars with the same F160W magnitude, is visible below $m_{\rm F160W} \sim 20.5$ and corresponds to the sequence $S_{2}$ identified in Fig.~\ref{fig:mappe}.  $S_{1}$ stars have intermediate $m_{\rm F606W}-m_{\rm F160W}$ and $m_{\rm F814W}-m_{\rm F160W}$ colors. Noticeably both $S_{1}$ and $S_{2}$ are spread over a relatively large interval of $m_{\rm F110W}-m_{\rm F160W}$.
  
Finally we remark that, as previously shown by King et al.\,(2012), the SGB is clearly bimodal in each CMD of Fig.~\ref{fig:cmdBMS}, with the bright SGB hosting the majority of SGB stars. Below we will show that the bright and faint SGBs are the progeny of MS-I and MS-II stars, respectively.

 In Fig.~\ref{fig:MSRLs} we have used red and blue colors to represent the MS-I and MS-II stars identified in Fig.~\ref{fig:cmdBMS} in three CMDs made by plotting the F814W magnitudes as a function of three different colors. Specifically, we use in the panel a1 the widest color baseline, $m_{\rm F606W}-m_{\rm F160W}$,  while in panel b1 and c1 we show the optical, $m_{\rm F606W}-m_{\rm F814W}$, and NIR, $m_{\rm F110W}-m_{\rm F160W}$, colors, respectively.
  We immediately note that the two main populations can be followed continuously from the turn-off towards the hydrogen burning limit, thus demonstrating that the groups of MS-I and MS-II stars identified in the distinct magnitude intervals of Fig.~\ref{fig:cmdBMS} actually trace the same stellar population along the entire CMD.
  The red and the blue lines over-imposed on each diagram are the fiducial lines of the two main populations and have been derived by using a procedure based on the naive estimator method (Silverman 1986) described in Sect.~\ref{subsec:bottom}.

 We also calculated the $m_{\rm F606W}-m_{\rm F160W}$ color difference, $\delta_{\rm F606W,F160W}$,  between the blue and the red fiducial lines and we show $m_{\rm F814W}$  against $\delta_{\rm F606W,F160W}$ in the panel a2. We obtained $\delta_{\rm F606W,F814W}$ and $\delta_{\rm F110W,F160W}$, similarly and plot $m_{\rm F814W}$ as a function of these quantities in panels b2 and c2.

 In the upper part of the CMD, MS-I is redder than MS-II in all the analyzed CMDs and the color separation increases when moving from the MSTO towards fainter magnitudes and has a maximum at $m_{\rm F814W} \sim 20.0$. Below that the MS-I and the MS-II get closer and merge together at the luminosity of the MS knee, around $m_{\rm F814W} \sim 21.0$.

   The MS-II is redder than the MS-I at fainter magnitudes in both the $m_{\rm F606W}-m_{\rm F160W}$ and $m_{\rm F110W}-m_{\rm F160W}$ color. 
   Above the MS knee, the separation between the two MSs is maximum in the $m_{\rm F606W}-m_{\rm F160W}$ color, where it approaches $\sim$0.17 mag, and decreases for shorter color baselines. At the bottom of the MS, the dependence of the MS separation from the color baseline is less evident and, in each of these two CMDs, it is close to 0.12 mag at $m_{\rm F814W} \sim 24.0$.
   In the optical CMD, the two MSs are nearly coincident in the magnitude interval $22.5 \lesssim m_{\rm F814W}\lesssim 24.0$ while MS-I is redder than MS-II at fainter luminosity.

 \section{Comparison with theory}\label{sec:teoria}
 To get information on the main populations of $\omega$\,Centauri we compare in 
Fig.~\ref{fig:iso} the observed CMDs with $\alpha$-enhanced ($[\alpha/Fe]=0.3$) isochrones from Dotter et al.\,(2008).
 The red isochrones correspond to a stellar population with [Fe/H]=$-$1.70 and primordial helium abundance Y=0.246, while the blue isochrone has [Fe/H]=$-$1.40 and Y=0.40. The adopted metallicities have been selected in order to match the observed CMD.
 We also show with green dashed lines the isochrones with Y=0.37 and [Fe/H]=$-1.70$. We have assumed a distance modulus, (m$-$M)$_{0}$=13.69 and a reddening E(B$-$V)=0.13 which are similar to the values listed by the Harris (1996, updated as in 2010) catalog, and an age of 13.5 Gyr. The parameters corresponding to the best-fitting isochrones have been determined as in Dotter et al.\,(2010).  Briefly, we started by setting the distance modulus, and reddening to the values provided by the 2010 version of the Harris\,(1996) catalog. We assumed for the MS-I  a metallicity, [Fe/H]=$-1.75$ corresponding to the peak of the metallicity distribution derived from high-resolution spectroscopy by Marino et al.\,(2011). 
    Moreover, we assumed that the MS-II is enhanced in [Fe/H] by 0.3 dex with respect to the MS-I $\omega$\,Centauri as inferred  Piotto et al.\,(2005). We have then adjusted the values of metallicity, reddening, and distance to optimize the fit with stars brighter than the MS knee. The age has been then estimated by selecting the isochrone providing the best match with the position and shape of the upper SGB.

 To better compare the relative colors of MS-I and MS-II stars with those of the isochrones we have calculated the color difference between each isochrone and the isochrone with [Fe/H]=$-1.70$ and primordial helium abundance.
 In the panels a2, b2, and c2 of Fig.~\ref{fig:iso}  we plot $m_{\rm F814W}$  against the color difference, $\delta$, derived from the isochrones (blue-continuous and green-dashed lines) and the corresponding color difference between MS-II and MS-I shown in Fig.~\ref{fig:MSRLs}.
 
We emphasize here that, according to studies based on high-resolution spectroscopy,  stars in $\omega$\,Centauri exhibit a complex metallicity distribution (e.g., Norris \& Da Costa 1995; Johnson \& Pilachowski 2010; Marino et al.\,2011). Therefore our choice of using only few isochrones is meant to reproduce the main features of the cluster CMD and the average properties of the two main populations of $\omega$\,Centauri and we refer to the paper by Tailo et al.\,(2016) for a more-complete comparison between the upper part of the observed CMD of $\omega$\,Centauri and theoretical isochrones.
 Nevertheless, we note that the value of [Fe/H] that we have adopted for the MS-I  roughly matches the main peak of the metallicity distribution derived by Marino et al.\,(2011) and the metallicity difference between the two isochrones is close to the average [Fe/H] difference of red-MS and blue-MS stars derived by Piotto et al.\,(2005).

 Figure~\ref{fig:iso} reveals that the three isochrones reproduce well the CMD region above the MS knee. Specifically, the metal-poor and helium-poor isochrone matches the MS-I and the helium-rich isochrones provide a good fit to the MS-II. 
 Noticeably, the helium content needed for the metal-poor isochrone to match the MS-II is Y=0.37, significantly lower than the helium abundance of the metal-rich isochrone (Y=0.40). The helium-rich population of $\omega$\,Centauri has been interpreted as a second stellar generation(s) formed by the ejecta of evolved intermediate-mass first-population stars (see e.g.\,D'Antona et al.\,2011, 2016). In particular, the presence of metal-rich stars with helium abundance of Y=0.40 is consistent with the scenario suggested by Karakas et al.\,(2014) where stars with extreme helium content formed from the ejecta of a previous generation of helium-rich stars.  

 The three isochrones qualitatively reproduce the shape of the two main SGBs, with the faint and the bright SGB corresponding to the helium-rich and helium-poor population, respectively. 
 These results are quite expected as the same conclusion has been obtained in several previous papers (e.g.\,Bedin et al.\,2004; Norris 2004; Piotto et al.\,2005; King et al.\,2012; Tailo et al.\,2016). In addition, we note that the MS-II is also consistent with a stellar population with the same metallicity as the MS-I but with different helium abundance.
 
 Below the MS knee the adopted isochrones overlap the MS-I and none of them reproduce the colors of MS-II stars. In particular, in the $m_{\rm F814W}$ vs.\,$m_{\rm F110W}-m_{\rm F160W}$ CMD the color separation between the metal-rich isochrones and the metal-poor ones is smaller than 0.03 mag for $m_{\rm F814W}>21.0$. 
 This fact indicates that, according to isochrones by Dotter et al.\,(2008), in the analyzed interval of He and Fe abundances the $m_{\rm F110W}-m_{\rm F160W}$ color of faint M-dwarfs is marginally affected by variations in helium and iron content.

    In this context, it is worth noting that variations in metallicity alone and variations in the C$+$N$+$O sum of $\omega$\,Centauri stars alone are not viable hypotheses to explain the observed behaviour of MS-I and MS-II. 
   The fact that the MS-II is more metal-rich than the MS-I was clearly demonstrated by Piotto et al.\,(2005) on the basis of direct spectroscopic measurements of iron abundance in bright MS-I and MS-II stars. Since the MS-I is redder than the MS-II above the MS knee, Piotto and collaborators noticed that the only way to reproduce the colors and the metallicity of the two MSs of $\omega$\,Centauri is to assume that the MS-II is also highly helium enhanced with respect to the MS-I. In this work we do not take into account the possibility that the MS-I stars have higher metallicity than MS-II and refer to papers by Piotto et al.\,(2005), Bellini et al.\,(2010), Tailo et al.\,(2016) and references therein for photometric and spectroscopic metallicity determination of MS-I and MS-II stars.
  Similarly, we note that difference in the overall C$+$N$+$O alone are not able to reproduce the difference in color between MS-I and MS-II stars that we measured in this paper. Indeed a C$+$N$+$O variation produces a marginal color difference for MS stars (e.g.\,Cassisi et al.\,2008; Ventura et al.\,2009; Sbordone et al.\,2011).

 \subsection{The effect of light elements on the colors of MS stars}
 \label{subsec:lightelements}
 \begin{centering} 
 \begin{figure*} 
   \includegraphics[height=8.5cm]{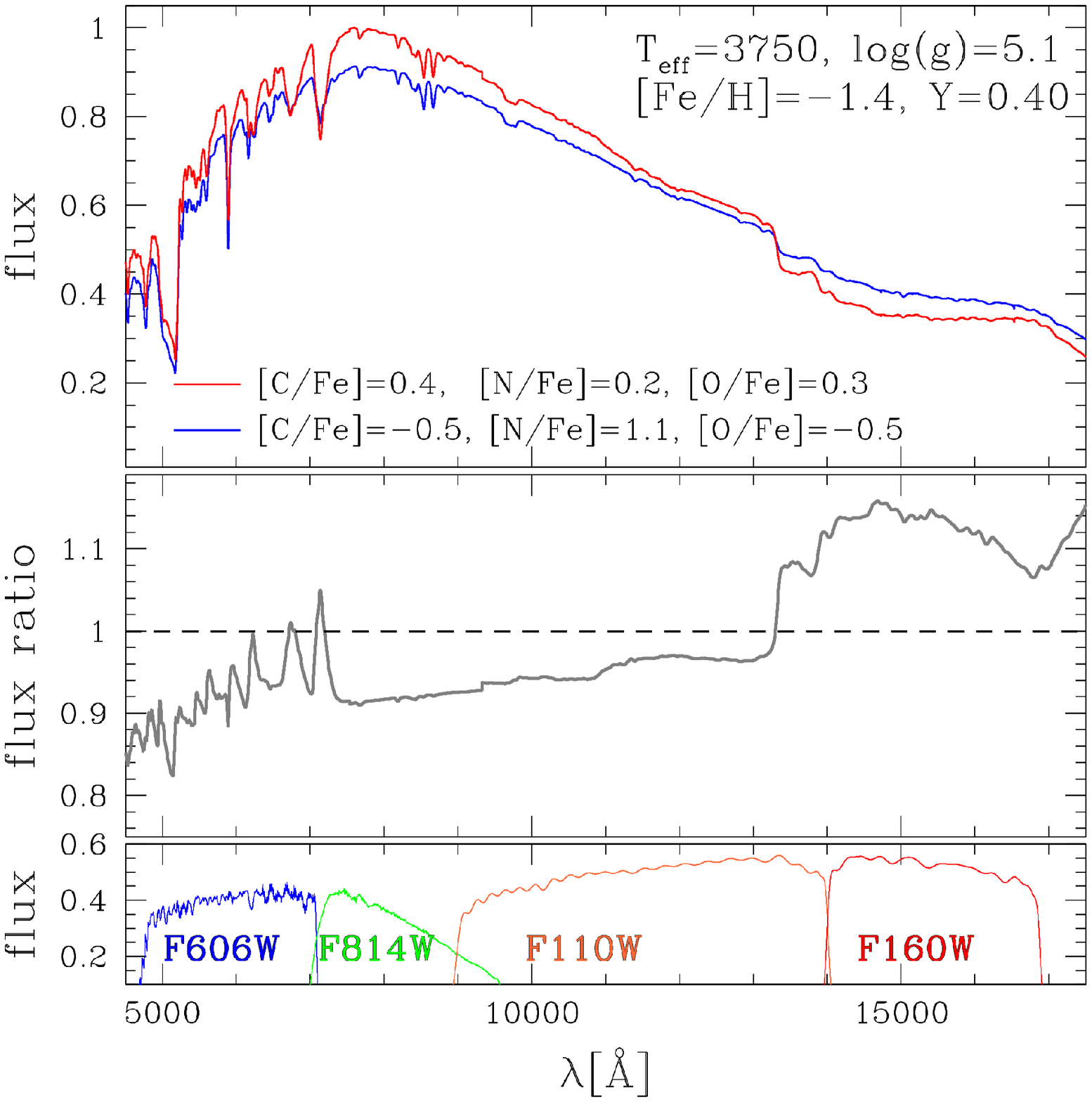}
   \includegraphics[height=8.5cm]{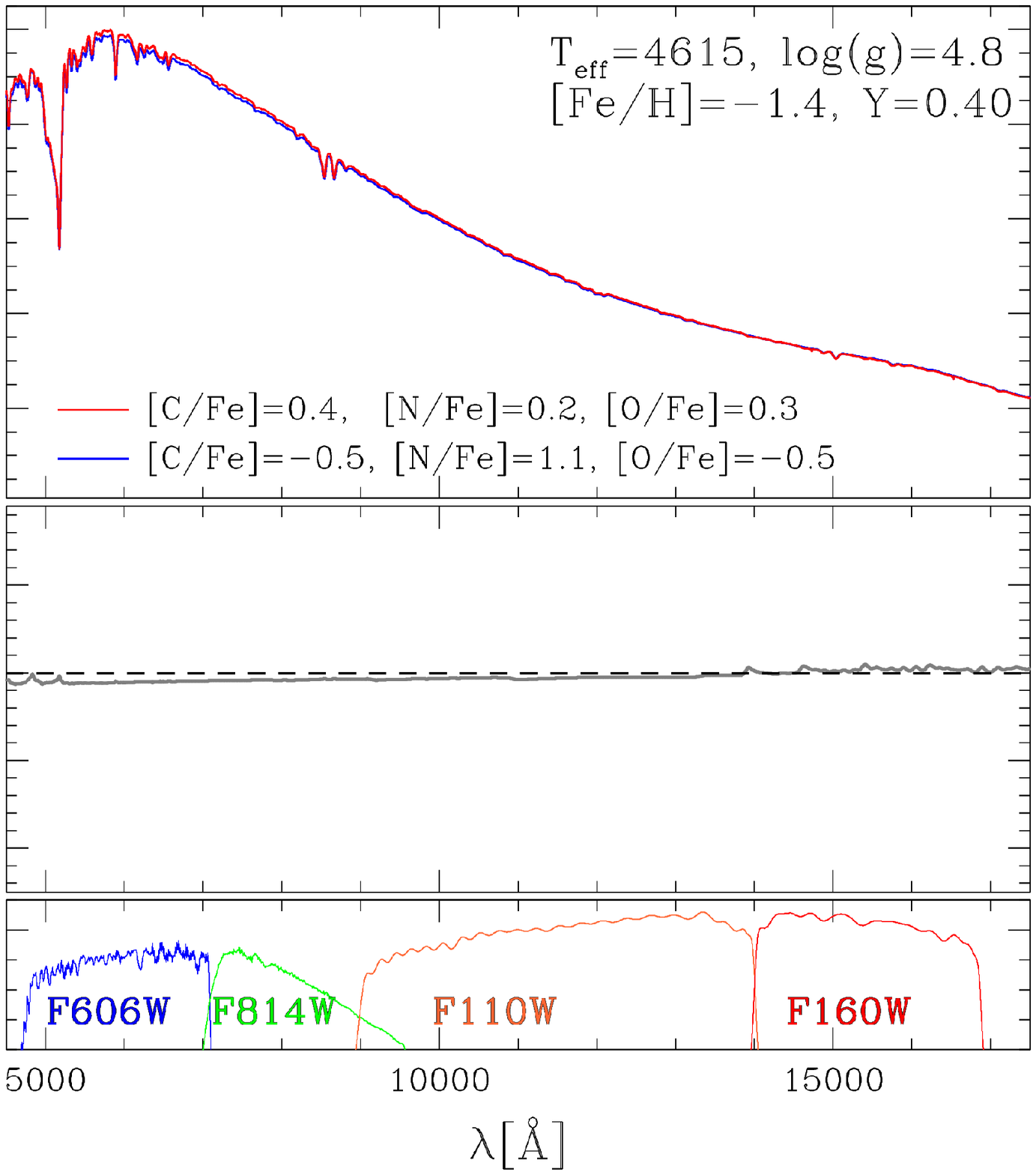}
   \caption{Synthetic spectra for a MS star with [Fe/H]=$-1.4$ and Y=0.40 located below (left panels) and above (right panels) the MS knee. Upper panels compare a blue and a red spectrum, named comparison and reference spectrum, respectively, with the same effective temperature and gravity but different chemical composition (see text for details). Middle panels show the ratio between the flux of the comparison and the reference spectrum as a function of the wavelength, while in lower panels we plot the transmission curves of the ACS/WFC and NIR/WFC3 spectra used in this paper.} 
  \label{fig:spettri} 
 \end{figure*} 
 \end{centering} 

To further investigate the physical reasons responsible for the multiple sequences,  we have extended the method introduced in our previous papers to $\omega$\,Centauri and compared the colors of the observed sequences with those predicted by appropriate synthetic spectra (see Milone et al.\,2012, 2014 for details).
Briefly, we first identified a series of reference points along the MS-I fiducial line with F814W magnitude ranging from 17.5 to 24.0 in steps of 0.5 mag. For each point we used the best-fitting isochrone of MS-I stars to derive the corresponding values of effective temperature ($T_{\rm eff}$) and gravity ($log{g}$). 
For each reference point we simulated two spectra with distinct C, N, O abundances. For the first spectrum (reference spectrum) we assumed [C/Fe]=0.4, [N/Fe]=0.2 and [O/Fe]=0.3, while for the second one (comparison spectrum) we adopted [C/Fe]=$-$0.5, [N/Fe]=1.1, and [O/Fe]=$-$0.5. The adopted C, N, O abundances match the abundances derived for these elements by Marino et al.\,(2011, 2012) and Johnson \& Pilachowski\,(2010).

These chemical abundances and values of $T_{\rm eff}$ and $log{g}$ have been adopted as input parameters  for the ATLAS12 and SYNTHE codes (Castelli 2005; Kurucz 2005; Sbordone et al.\,2007) to generate a grid of synthetic spectra in the wavelength range between 4500 and 17500 \AA. We have included the following molecules in the Kurucz compilation: CO, C$_{2}$, CN, OH, MgH, SiH, H$_{2}$O, TiO, VO, ZrO (Partridge \& Schwenke\,1997; Schwenke\,1998; B.\,Plez, private communication).
The resulting spectra for two stars below and above the knee, are illustrated in Fig.~\ref{fig:spettri}, and correspond to MS stars with $m_{\rm F814W}=23.25$ and 20.75, respectively. Upper panels compare the reference and the comparison spectra while lower panels show the flux ratio between the two spectra as a function of the wavelength. We note that low-temperature spectra with different chemical composition are very different along most of the analyzed wavelength range, whereas the two hot spectra are very similar.  In particular, we confirm previous findings that the strong flux difference of M dwarfs in the spectral region with $\lambda \gtrsim 14,000 \AA$ is mostly due to oxygen  abundance in the atmospheres of MS-I and MS-II stars through the absorption of the H$_{2}$O molecules (Milone et al.\,2012; 2014; Dotter et al.\,2015).

 The synthetic spectra have been integrated over the transmission curves of the ACS/WFC and WFC3/NIR filters used in this paper to derive synthetic magnitudes. For each reference point we have then calculated the difference between the F606W, F814W, F110W, and F160W magnitudes obtained from the comparison and the reference spectrum (d$mag$). 

 We confirm the previous finding that, above the knee, the effect of C, N, O variations is negligible when both the optical and NIR colors are used (e.g.\,Sbordone et al.\,2011; Milone 2015; Dotter et al.\,2015). In contrast, the effect of light elements strongly affects the flux of fainter than the knee stars in the optical and near-infrared bands used in this paper.

\begin{centering} 
 \begin{figure*} 
   \includegraphics[height=8.5cm]{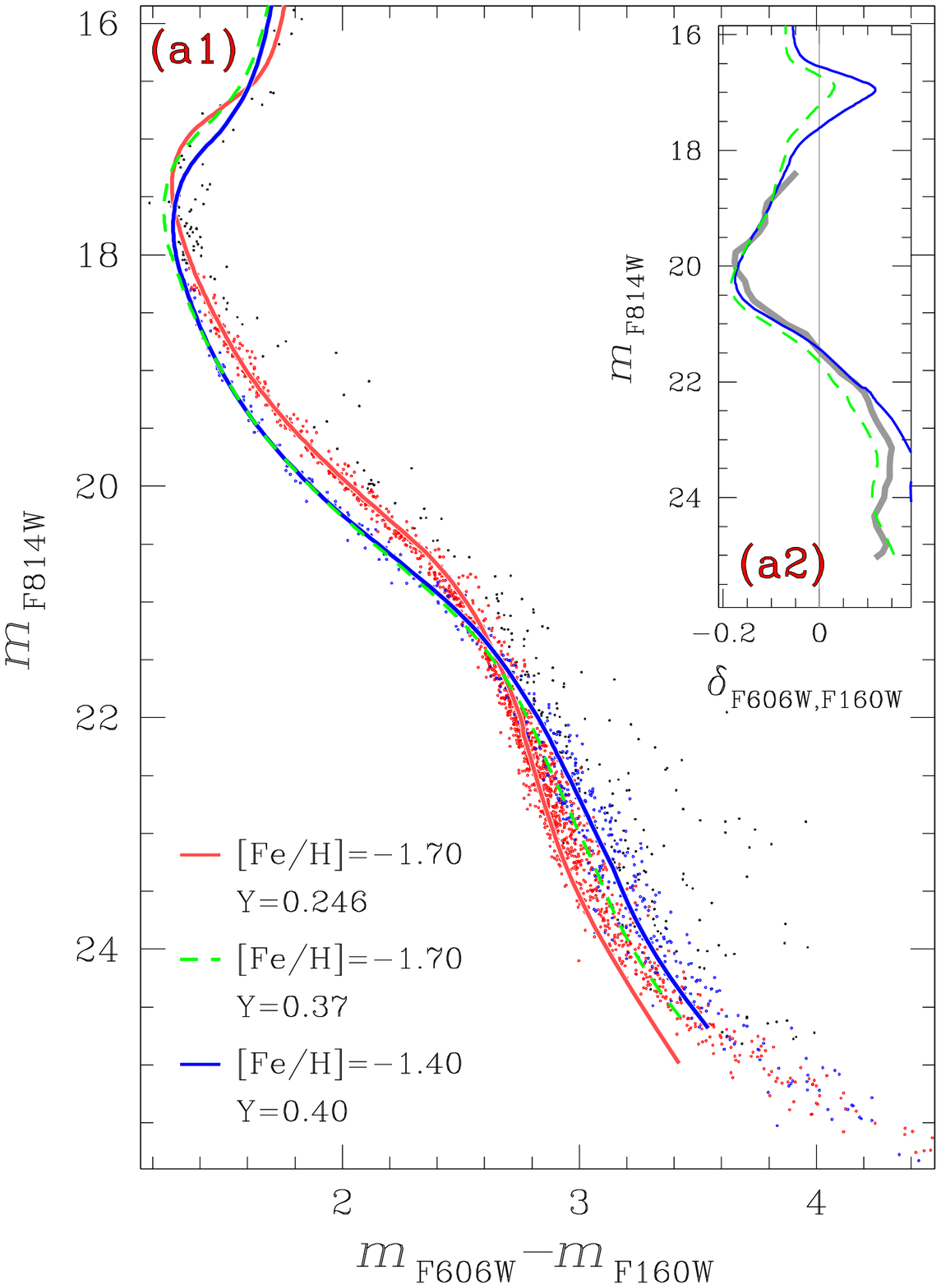}
   \includegraphics[height=8.5cm]{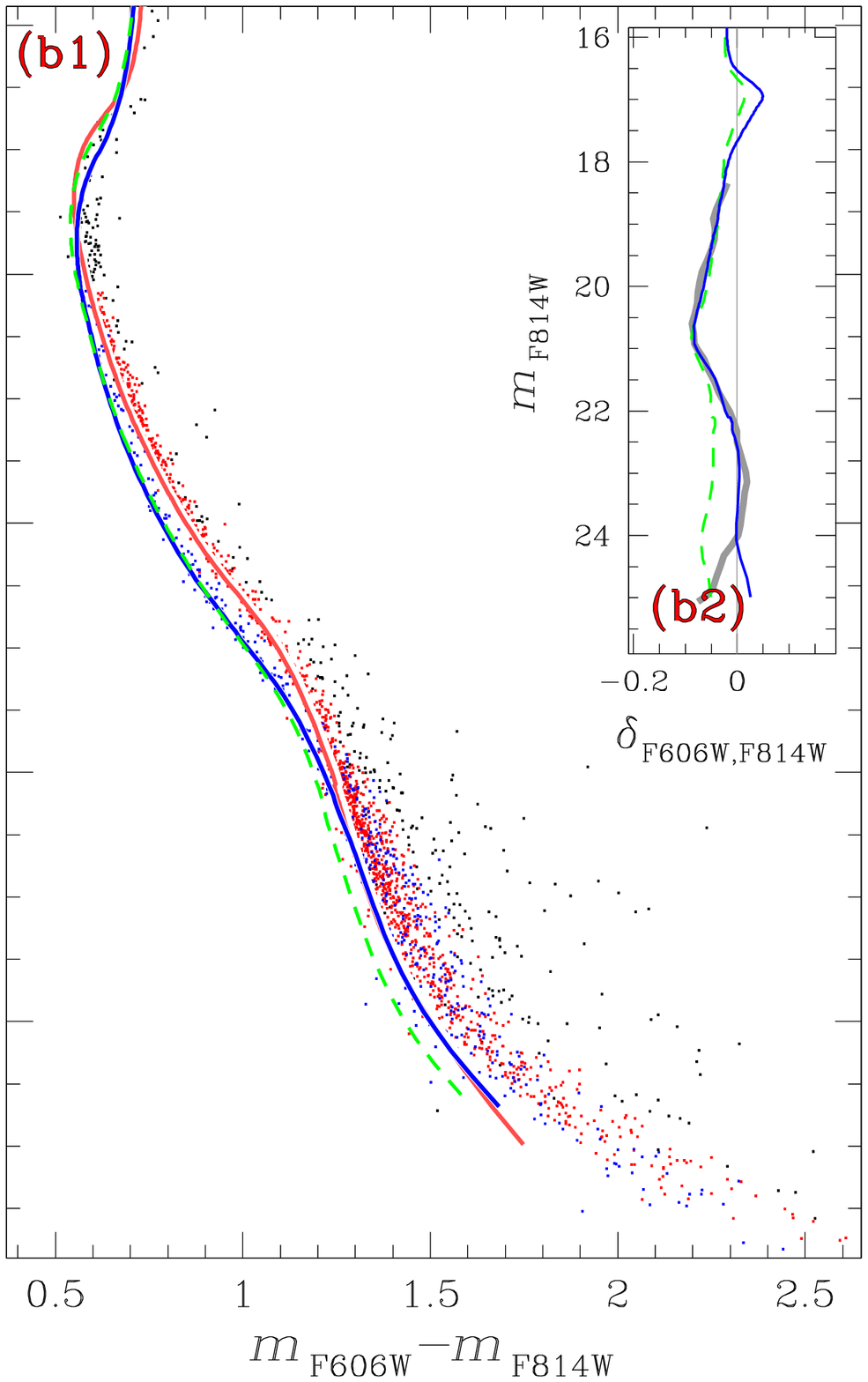}
   \includegraphics[height=8.5cm]{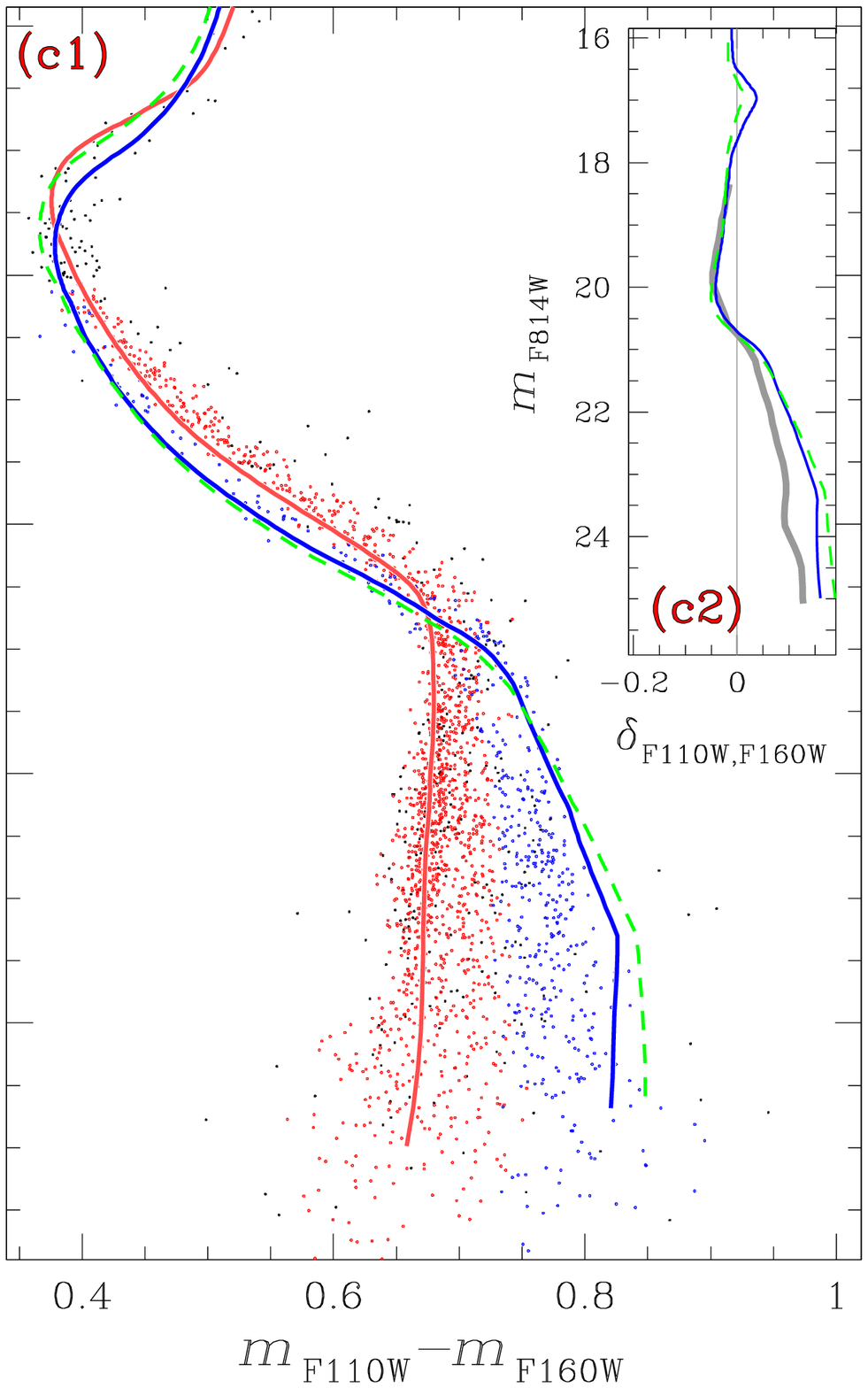} 
   \caption{As in Fig.~\ref{fig:iso} but in this case, the blue and green isochrones account for the variation of C, N, O typical of second-population stars in $\omega$\,Centauri (see text for details).} 
  \label{fig:iso2} 
 \end{figure*} 
 \end{centering}

 The blue and green isochrones plotted in Fig.~\ref{fig:iso2} have been obtained by adding to the helium-rich isochrones the corresponding values of d$mag$. These isochrones provide a  qualitatively better fit of the entire CMD, from the SGB to the bottom of the explored MS.   

\subsection{The stellar subpopulations of $\omega$\,Centauri}
 The six subpopulations identified in Sect.~\ref{subsec:bottom} are visible over a small magnitude interval below the MS knee. To constrain their main properties we combine the observations of this paper with isochrones from Dotter et al.\,(2008) and literature results.

\begin{centering} 
 \begin{figure*} 
   \includegraphics[height=12.0cm]{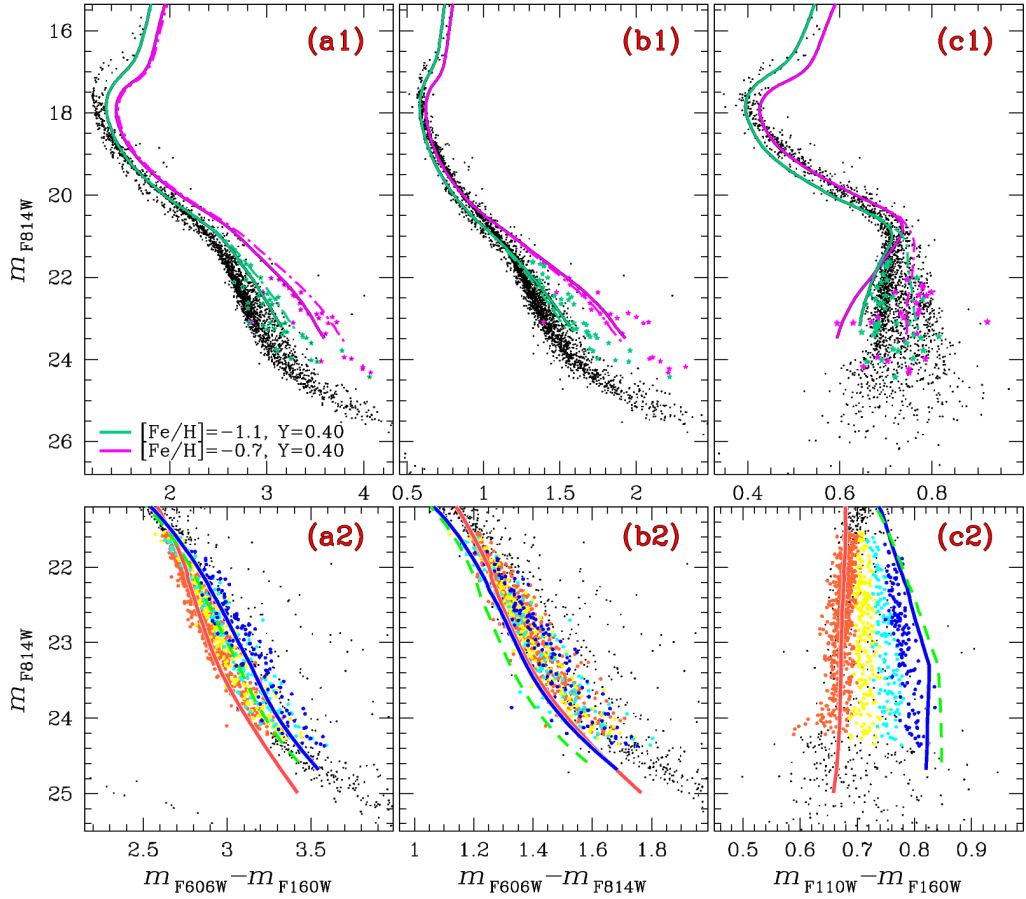}
   \caption{{\textit{Upper Panels.} Reproduction of the CMDs of Fig.~\ref{fig:MSRLs} with $S_{1}$ and $S_{2}$ stars colored aqua and magenta, respectively. The aqua and magenta continuous lines are helium-rich isochrones with [Fe/H]=$-1.0$ and [Fe/H]=$-0.7$, respectively. The corresponding dashed-dotted isochrones account for the C, N, O abundance of $S_{1}$ and $S_{2}$ stars with extreme chemical composition. \textit{Lower Panels.} Zoom-in of the CMDs shown in the upper panels on the lower MS. Stars in the sub-populations A, B, C, and D are colored orange, yellow, cyan, and blue, respectively. The isochrones introduced in Fig.~\ref{fig:iso2} are overimposed on the CMDs. All the isochrones plotted in this figure have ages of 13.5 Gyr.}} 
  \label{fig:iso3} 
 \end{figure*} 
 \end{centering}

            In the upper panels of Fig.~\ref{fig:iso3} we reproduce the three CMDs of Fig.~\ref{fig:MSRLs} and mark $S_{1}$ and $S_{2}$ stars with aqua and magenta starred symbols, respectively. We note that both subpopulations $S_{1}$ and $S_{2}$ have larger $m_{\rm F606W}-m_{\rm F160W}$ and $m_{\rm F606W}-m_{\rm F814W}$ colors than the majority of MS stars with the same $m_{\rm F814W}$ luminosity. In contrast, $S_{1}$ and $S_{2}$ stars are mostly overimposed to the remaining MS stars in the $m_{\rm F814W}$ vs.\,$m_{\rm F110W}-m_{\rm F160W}$ CMD. On average, $S_{2}$ stars are redder than $S_{1}$ stars with similar F814W magnitude level in all the CMDs of Fig.~\ref{fig:iso3}.
            
            High-resolution spectroscopy reveals that $\omega$\,Centauri hosts stellar populations which are highly iron enhanced with respect to the majority of $\omega$\,Centauri stars ([Fe/H]$\gtrsim-1.4$, e.g.\,Norris \& Da Costa 1995; Johnson et al.\,2009). The most-metal rich stars have [Fe/H]=$\sim -0.7$ (Marino et al.\,2011) and correspond to the MS-a discovered by Bedin et al.\,(2004).
 Further information on MSa stars come from the study based on multi-wavelength  {\it HST} photometry by Bellini et al.\,(2010), who show that MSa stars are almost overimposed to the red MS in the $m_{\rm F814W}$ vs.\,$m_{\rm F606W}-m_{\rm F814W}$ plane. This fact demonstrates that MSa stars are highly helium enhanced with respect to the majority of stars in $\omega$\,Centauri.

            The magenta continuous isochrones overimposed to the CMDs of Fig.~\ref{fig:iso3} have metallicity, $\alpha$-elements and helium abundances ([$\alpha$/Fe]=0.4, [Fe/H]=$-0.7$, and Y=0.40) consistent with those inferred from the spectroscopic and photometric papers by Marino et al.\,(2011) and Bellini et al.\,(2010).
            The magenta dashed-dotted isochrones have the same chemical composition as the continuous ones but are depleted in oxygen by 0.5 dex and are derived as described in Sect.~\ref{subsec:lightelements}.            

            The aqua isochrones correspond to stellar populations with [Fe/H]=$-1.1$ and [$\alpha$/Fe]=0.4 but different relative abundance of C, N, and O. In this case, the dashed-dotted isochrones are enhanced in N by 1.0 dex and depleted in carbon and oxygen by 0.6 and 0.4 dex with respect to the continuous isochrones. The adopted values of C, N, and O are consistent with those derived by Marino et al.\,(2012) from high-resolution spectroscopy.   

            The adopted isochrones qualitatively match the observed colors and magnitudes of stars in the upper panels of Fig.~\ref{fig:iso3}. In particular, the fact that $S_{1}$ and $S_{2}$ stars are distributed between the continuous and the dashed-dotted isochrones in the $m_{\rm F814W}$ vs.\,$m_{\rm F110W}-m_{\rm F160W}$ CMD suggests that the most metal-rich populations of $\omega$\,Centauri exhibit star-to-star oxygen variations.

            The subpopulation of MS-I and MS-II are analyzed 
             in the lower panels of Fig.~\ref{fig:iso3} where we use  orange, yellow, cyan, and blue colors to represent the stars of the four populations, A--D, identified in Sect.~\ref{subsec:bottom}. We also overimpose to each CMD the isochrones introduced in Fig.~\ref{fig:iso2}.
            The four stellar sub-populations are well mixed in the $m_{\rm F814W}$ vs.\,$m_{\rm F606W}-m_{\rm F814W}$ CMD, but have different $m_{\rm F606W}-m_{\rm F160W}$ and $m_{\rm F110W}-m_{\rm F160W}$ colors.

            As discussed in previous papers (e.g.\,Milone et al.\,2012, 2014; Dotter et al.\,2014) and in Sect.~\ref{subsec:lightelements}, the $m_{\rm F110W}-m_{\rm F160W}$ color is poorly sensitive to the stellar metallicity and the helium abundance but is strongly affected by the oxygen abundance. The fact that the sub-populations A and B of MS-I exhibit, on average, different $m_{\rm F110W}-m_{\rm F160W}$ colors suggests that their stars mostly differ in their light-element abundance. Similar conclusions can be extended to  the sub-populations C and D of the MS-II.

\section{Summary and conclusions}\label{sec:conclusioni}
We have presented a study of multiple stellar populations in $\omega$\,Centauri  based on multi-epoch and deep images collected with the WFC/ACS and NIR/WFC3 cameras of {\it HST} as part of the {\it HST} large program on $\omega$\,Centauri (GO-14118$+$14622, PI.\,L.\,R.\,Bedin).  We have used the method based on effective-PSF fitting (e.g.\,Anderson \& King 2003; Anderson et al.\,2008) to derive high-precision stellar photometry and astrometry and determined stellar proper motions to separate cluster members from field stars.
 
The cluster CMD reveals multiple sequences over a range of at least eight magnitudes in the F814W band. By using CMDs and pseudo-CMDs made with appropriate combinations of magnitudes, we have identified the two principal MSs (MS-I, and MS-II) of $\omega$\,Centauri and for the first time we have followed them continuously, from the MS turn off to the bottom of the MS.
To investigate multiple populations among M-dwarfs we have adapted to faint MS stars  the method used by Milone et al.\,(2015, 2017) and we have derived pseudo two-color diagrams, or -- by using the nomenclature of our previous papers -- chromosome maps, by using NIR and optical colors.
  In this case, the pseudo-color, $\Delta_{2}$, plotted y-axis of the chromosome map is derived from the $m_{\rm F110W}-m_{\rm F160W}$ color and is mostly sensitive to stellar populations with different oxygen abundance. The quantity plotted on the x axis, $\Delta_{1}$, is made by combining F606W, F814W, and F160W magnitudes and it is also sensitive to the metallicity of the stellar populations.

 We have discovered that below the MS knee, both MS-I and MS-II host stellar  subpopulations and we have identified  at least four stellar groups A, B, C, and D. Two additional stellar sequences ($S_{1}$ and $S_{2}$) are visible both in the chromosome map and on the red side of the most-populated MSs in the $m_{\rm F814W}$ vs.\,$m_{\rm F606W}-m_{\rm F160W}$ CMD.  They are consistent with $\alpha$-enhanced isochrones with Y=0.40 and [Fe/H]=$-1.1$ and $-0.7$, respectively, and each of them is not consistent with a simple stellar population. 
 
 In CMDs made with optical ($m_{\rm F606W}-m_{\rm F814W}$), NIR ($m_{\rm F160W}$ vs.\,$m_{\rm F110W}-m_{\rm F160W}$) and mixed ($m_{\rm F606W}-m_{\rm F160W}$) colors,  the MS-I is redder than the MS-II in the upper part of the CMD and the two MSs merge together at the luminosity of the MS knee.
In the NIR color, the MS-I and the MS-II are well separated below the MS knee but their color order is inverted, with the MS-I being redder than the MS-II.
 A similar MSs pattern has been previously observed in the massive GC NGC\,2808, where the two helium-rich MSs behave like the MS-II of $\omega$\,Centauri, while the helium-normal MS would correspond to the MS-I (Milone et al.\,2012).  
 The CMD of M-dwarfs in $\omega$\,Centauri is more complex than that of previously-studied GCs (e.g.\,Milone et al.\,2012, 2014; Richer et al.\,2013; Correnti et al.\,2016) and hosts at least six stellar sub-populations.

 We have compared the observed CMDs with appropriate isochrones where we have accounted for the effect of metallicity, helium, and C, N, O on the stellar colors. We confirm that the MS-I corresponds to a metal-poor stellar population
 ([Fe/H]$\sim -1.7$) with primordial helium abundance and high oxygen content ([O/Fe]$\sim$0.4). MS-II is well fitted both by a metal-poor and a metal-rich isochrone with [Fe/H]=$-1.7$ and [Fe/H]=$-1.4$. In order to reproduce the MS region above the knee, both isochrones should be strongly-enhanced in helium, with the metal-poor and metal-rich isochrone having Y=0.37 and Y=0.40, respectively. These results suggest that MS-II hosts helium-rich stars that span a wide range of metallicity.

\section*{acknowledgments} 
\small
We are gratefull to the anonymous referee for comments and suggestions that have improved the quality of this manuscript.
APM and AFM acknowledge support by the Australian Research Council through Discovery Early Career Researcher Awards DE150101816 and DE160100851.
The USA authors acknowledge the support for program number GO-14118 and GO-14662 provided by NASA through a grant from the Space Telescope Science Institute.

\bibliographystyle{aa}

\begin{thebibliography}{}

\bibitem[Anderson(1997)]{1997PhDT.........8A} Anderson, A.~J.\ 1997, Ph.D.~Thesis, 1153 
  
\bibitem[Anderson \& King(2000)]{2000PASP..112.1360A} Anderson, J., \& King, I.~R.\ 2000, \pasp, 112, 1360 

\bibitem[Anderson \& King(2003)]{2003AJ....126..772A} Anderson, J., \& King, I.~R.\ 2003, \aj, 126, 772 
  
\bibitem[Anderson et al.(2008)]{2008AJ....135.2055A} Anderson, J., Sarajedini, A., Bedin, L.~R., et al.\ 2008, \aj, 135, 2055 

\bibitem[Bedin et al.(2003)]{2003AJ....126..247B} Bedin, L.~R., Piotto, G., King, I.~R., \& Anderson, J.\ 2003, \aj, 126, 247 
  
\bibitem[Bedin et al.(2004)]{2004ApJ...605L.125B} Bedin, L.~R., Piotto, G., Anderson, J., et al.\ 2004, \apjl, 605, L125 

\bibitem[Bedin et al.(2005)]{2005MNRAS.357.1038B} Bedin, L.~R., Cassisi, S., Castelli, F., et al.\ 2005, \mnras, 357, 1038 

\bibitem[Bedin et al.(2009)]{2009ApJ...697..965B} Bedin, L.~R., Salaris, M., Piotto, G., et al.\ 2009, \apj, 697, 965 
  
\bibitem[Bellini et al.(2010)]{2010AJ....140..631B} Bellini, A., Bedin, L.~R., Piotto, G., et al.\ 2010, \aj, 140, 631 

\bibitem[Bellini et al.(2013)]{2013ApJ...769L..32B} Bellini, A., Anderson, J., Salaris, M., et al.\ 2013, \apjl, 769, L32 

\bibitem[Brown \& Wallerstein(1993)]{1993AJ....106..133B} Brown, J.~A., \& Wallerstein, G.\ 1993, \aj, 106, 133 

\bibitem[Cassisi et al.(2008)]{2008ApJ...672L.115C} Cassisi, S., Salaris, M., Pietrinferni, A., et al.\ 2008, \apjl, 672, L115 
    
\bibitem[Castelli(2005)]{2005MSAIS...8...25C} Castelli, F.\ 2005, Memorie della Societa Astronomica Italiana Supplementi, 8, 25   
  
\bibitem[Correnti et al.(2016)]{2016ApJ...823...18C} Correnti, M., Gennaro, M., Kalirai, J.~S., Brown, T.~M., \& Calamida, A.\ 2016, \apj, 823, 18 

\bibitem[D'Antona et al.(2011)]{2011ApJ...736....5D} D'Antona, F., D'Ercole, A., Marino, A.~F., et al.\ 2011, \apj, 736, 5 

\bibitem[D'Antona et al.(2016)]{2016MNRAS.458.2122D} D'Antona, F., Vesperini, E., D'Ercole, A., et al.\ 2016, \mnras, 458, 2122 
  
\bibitem[Dotter et al.(2008)]{2008ApJS..178...89D} Dotter, A., Chaboyer, B., Jevremovi{\'c}, D., et al.\ 2008, \apjs, 178, 89-101 

\bibitem[Dotter et al.(2010)]{2010ApJ...708..698D} Dotter, A., Sarajedini, A., Anderson, J., et al.\ 2010, \apj, 708, 698 
  
\bibitem[Dotter et al.(2015)]{2015MNRAS.446.1641D} Dotter, A., Ferguson, J.~W., Conroy, C., et al.\ 2015, \mnras, 446, 1641 
  
\bibitem[Gilliland(2004)]{2004acs..rept...17G} Gilliland, R.~L.\ 2004, Instrument Science Report ACS 2004-01, 18 pages,  

\bibitem[Harris(1996)]{1996AJ....112.1487H} Harris, W.~E.\ 1996, \aj, 112, 1487   

\bibitem[Johnson et al.(2008)]{2008ApJ...681.1505J} Johnson, C.~I., Pilachowski, C.~A., Simmerer, J., \& Schwenk, D.\ 2008, \apj, 681, 1505-1523 
    
\bibitem[Johnson et al.(2009)]{2009ApJ...698.2048J} Johnson, C.~I., Pilachowski, C.~A., Michael Rich, R., \& Fulbright, J.~P.\ 2009, \apj, 698, 2048 
  
\bibitem[Johnson \& Pilachowski(2010)]{2010ApJ...722.1373J} Johnson, C.~I., \& Pilachowski, C.~A.\ 2010, \apj, 722, 1373 

\bibitem[Karakas et al.(2014)]{2014ApJ...784...32K} Karakas, A.~I., Marino, A.~F., \& Nataf, D.~M.\ 2014, \apj, 784, 32 
  
\bibitem[King et al.(2012)]{2012AJ....144....5K} King, I.~R., Bedin, L.~R., Cassisi, S., et al.\ 2012, \aj, 144, 5 

\bibitem[Kurucz(2005)]{2005MSAIS...8...14K} Kurucz, R.~L.\ 2005, Memorie della Societa Astronomica Italiana Supplementi, 8, 14   
  
\bibitem[Lee et al.(1999)]{1999Natur.402...55L} Lee, Y.-W., Joo, J.-M., Sohn, Y.-J., et al.\ 1999, \nat, 402, 55 

\bibitem[Lindegren et al.(2016)]{2016A&A...595A...4L} Lindegren, L., Lammers, U., Bastian, U., et al.\ 2016, \aap, 595, A4   

\bibitem[Marino et al.(2010)]{2010IAUS..268..183M} Marino, A.~F., Piotto, G., Gratton, R., et al.\ 2010, IAU Symposium, 268, 183 

\bibitem[Marino et al.(2011)]{2011ApJ...731...64M} Marino, A.~F., Milone, A.~P., Piotto, G., et al.\ 2011, \apj, 731, 64 

\bibitem[Marino et al.(2012)]{2012ApJ...746...14M} Marino, A.~F., Milone, A.~P., Piotto, G., et al.\ 2012, \apj, 746, 14 

\bibitem[Milone et al.(2009)]{2009A&A...497..755M} Milone, A.~P., Bedin, L.~R., Piotto, G., \& Anderson, J.\ 2009, \aap, 497, 755 

\bibitem[Milone et al.(2012)]{2012A&A...540A..16M} Milone, A.~P., Piotto, G., Bedin, L.~R., et al.\ 2012, \aap, 540, A16

\bibitem[Milone et al.(2012)]{2012ApJ...754L..34M} Milone, A.~P., Marino, A.~F., Cassisi, S., et al.\ 2012, \apjl, 754, L34 

\bibitem[Milone et al.(2014)]{2014MNRAS.439.1588M} Milone, A.~P., Marino, A.~F., Bedin, L.~R., et al.\ 2014, \mnras, 439, 1588 

\bibitem[Milone(2015)]{2015MNRAS.446.1672M} Milone, A.~P.\ 2015, \mnras, 446, 1672 
  
\bibitem[Milone et al.(2015)]{2015MNRAS.447..927M} Milone, A.~P., Marino, A.~F., Piotto, G., et al.\ 2015, \mnras, 447, 927 

\bibitem[Milone et al.(2017)]{2017MNRAS.464.3636M} Milone, A.~P., Piotto, G., Renzini, A., et al.\ 2017, \mnras, 464, 3636 
  
\bibitem[Norris \& Da Costa(1995)]{1995ApJ...447..680N} Norris, J.~E., \& Da Costa, G.~S.\ 1995, \apj, 447, 680 

\bibitem[Norris(2004)]{2004ApJ...612L..25N} Norris, J.~E.\ 2004, \apjl, 612, L25 

\bibitem[Pancino et al.(2000)]{2000ApJ...534L..83P} Pancino, E., Ferraro, F.~R., Bellazzini, M., Piotto, G., \& Zoccali, M.\ 2000, \apjl, 534, L83 

\bibitem[Pancino et al.(2002)]{2002ApJ...568L.101P} Pancino, E., Pasquini, L., Hill, V., Ferraro, F.~R., \& Bellazzini, M.\ 2002, \apjl, 568, L101 
  
\bibitem[Partridge \& Schwenke(1997)]{1997JChPh.106.4618P} Partridge, H., \& Schwenke, D.~W.\ 1997, J.\,Chem.\,Phys., 106, 4618 
  
\bibitem[Piotto et al.(2005)]{2005ApJ...621..777P} Piotto, G., Villanova, S., Bedin, L.~R., et al.\ 2005, \apj, 621, 777 

\bibitem[Renzini et al.(2015)]{2015MNRAS.454.4197R} Renzini, A., D'Antona, F., Cassisi, S., et al.\ 2015, \mnras, 454, 4197   
  
\bibitem[Richer et al.(2013)]{2013ApJ...771L..15R} Richer, H.~B., Heyl, J., Anderson, J., et al.\ 2013, \apjl, 771, L15 
  
\bibitem[Sbordone et al.(2007)]{2007IAUS..239...71S} Sbordone, L., Bonifacio, P., \& Castelli, F.\ 2007, Convection in Astrophysics, 239, 71 

\bibitem[Sbordone et al.(2011)]{2011A&A...534A...9S} Sbordone, L., Salaris, M., Weiss, A., \& Cassisi, S.\ 2011, \aap, 534, A9   

\bibitem[Schwenke(1998)]{1998FaDi..109..321S} Schwenke, D.~W.\ 1998, Faraday Discussions, 109, 321 
  
\bibitem[Silverman(1986)]{1986desd.book.....S} Silverman, B.~W.\ 1986, Monographs on Statistics and Applied Probability, London: Chapman and Hall, 1986,  

\bibitem[Sollima et al.(2005)]{2005ApJ...634..332S} Sollima, A., Pancino, E., Ferraro, F.~R., et al.\ 2005, \apj, 634, 332 

\bibitem[Stanford et al.(2010)]{2010ApJ...714.1001S} Stanford, L.~M., Da Costa, G.~S., \& Norris, J.~E.\ 2010, \apj, 714, 1001 
  
\bibitem[Tailo et al.(2016)]{2016MNRAS.457.4525T} Tailo, M., Di Criscienzo, M., D'Antona, F., Caloi, V., \& Ventura, P.\ 2016, \mnras, 457, 4525   

\bibitem[Ventura et al.(2009)]{2009MNRAS.399..934V} Ventura, P., Caloi, V., D'Antona, F., et al.\ 2009, \mnras, 399, 934 
  
\bibitem[Villanova et al.(2007)]{2007ApJ...663..296V} Villanova, S., Piotto, G., King, I.~R., et al.\ 2007, \apj, 663, 296 

\bibitem[Villanova et al.(2014)]{2014ApJ...791..107V} Villanova, S., Geisler, D., Gratton, R.~G., \& Cassisi, S.\ 2014, \apj, 791, 107
  
\end{thebibliography}

\end{document}